%% file: main.tex
\documentclass[acmsmall,nonacm]{acmart}

\setcopyright{none}                
\renewcommand\footnotetextcopyrightpermission[1]{}

\AtBeginDocument{%
  }

\usepackage{titletoc}

\titlecontents{section}
  [1.5em] 
  {\addvspace{4pt}\bfseries} 
  {\contentslabel{1.5em}} 
  {\hspace*{-1.5em}} 
  {\titlerule*[0.4pc]{.}\contentspage} 

\titlecontents{subsection}
  [3.3em] 
  {\addvspace{2pt}} 
  {\contentslabel{2.3em}} 
  {\hspace*{-2.3em}} 
  {\titlerule*[0.4pc]{.}\contentspage} 

\titlecontents{subsubsection}
  [5.5em] 
  {\addvspace{1pt}} 
  {\contentslabel{3.0em}} 
  {\hspace*{-3.0em}} 
  {\titlerule*[0.4pc]{.}\contentspage} 

\usepackage{graphicx}
\usepackage{subcaption}   
\usepackage{caption}
\usepackage{comment}
\usepackage{todonotes}
\usepackage{xcolor}
\usepackage{etoolbox}
\usepackage{siunitx}

\usepackage{amsmath,amssymb}
\usepackage{arydshln}
\usepackage{multirow} 
\usepackage{enumitem} 
\usepackage{marvosym} 
\usepackage{booktabs}  
\usepackage{makecell}  

\usepackage{pifont} 
\usepackage{tikz}

\makeatletter
\patchcmd{\@mkauthors@i}
  {\MakeUppercase}
  {}%
  {}{\PackageWarning{acmart}{Failed to patch \\@mkauthors@i, author names may still be uppercase}}
\makeatother

\usepackage{xspace}
\newcommand{\sysname}{RelayGR\xspace}

\makeatletter
\patchcmd{\@mktitle@i}{\raggedright}{\centering}{}%
  {\PackageWarning{acmart}{Failed to patch \@mktitle@i (title may remain left-aligned)}}
\makeatother

\begin{document}

\begin{tikzpicture}[remember picture,overlay]
  \node[anchor=north west, xshift=1.35cm, yshift=-0.6cm]
       at (current page.north west)
       {\includegraphics[height=1.2cm]{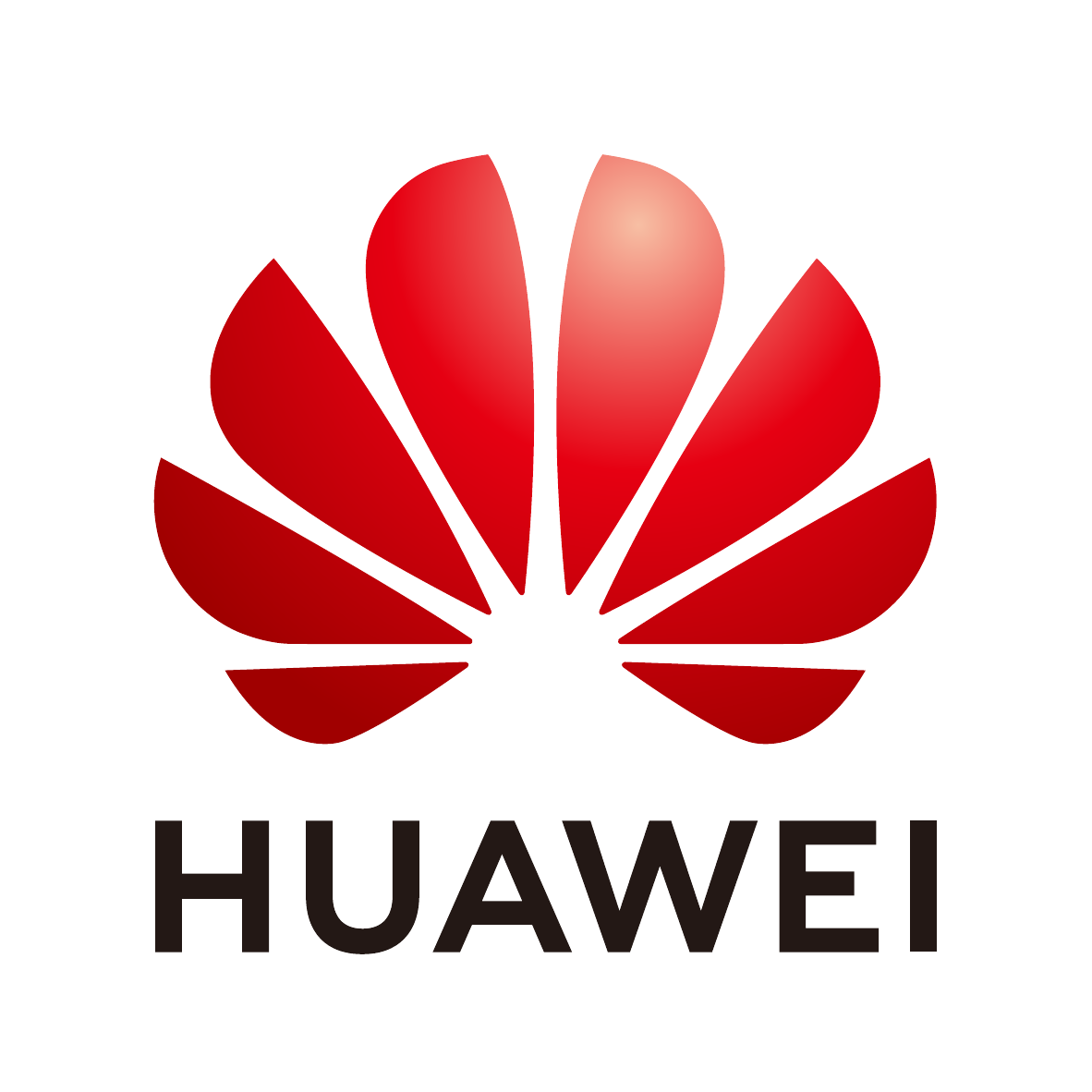}};
\end{tikzpicture}

\vspace{-1.7cm}
\begin{center}
  \rule{1\linewidth}{1pt} 
\end{center}
\vspace{0.4cm}
\title{RelayGR: Scaling Long-Sequence Generative Recommendation via Cross-Stage Relay-Race Inference}

\author{
    \centering
    Jiarui~Wang\textsuperscript{*}, Huichao~Chai\textsuperscript{*}, Yuanhang~Zhang, Zongjin~Zhou, Wei~Guo, Xingkun~Yang, Qiang~Tang, Bo~Pan, Jiawei~Zhu, Ke~Cheng\textsuperscript{\dag}, Yuting~Yan\textsuperscript{\dag}, Shulan~Wang\textsuperscript{\dag}, Yingjie~Zhu\textsuperscript{\dag}, Zhengfan~Yuan, Jiaqi~Huang, Yuhan~Zhang, Xiaosong~Sun, Zhinan~Zhang, Hong~Zhu, Yongsheng~Zhang, Tiantian~Dong, Zhong~Xiao, Deliang~Liu, Chengzhou~Lu, Yuan~Sun, Zhiyuan~Chen, Xinming~Han, Zaizhu~Liu, Yaoyuan~Wang, Ziyang~Zhang, Yong~Liu, Jinxin~Xu, Yajing~Sun, Zhoujun~Yu, Wenting~Zhou, Qidong~Zhang, Zhengyong~Zhang, Zhonghai~Gu, Yibo~Jin\textsuperscript{\Letter}, Yongxiang~Feng\textsuperscript{\Letter}, Pengfei~Zuo\textsuperscript{\Letter} \\
    \vspace{0.3cm}
    \textit{Huawei Technologies Co., Ltd.}\\
}

\thanks{\textsuperscript{*}Equal contribution.}
\thanks{\textsuperscript{\dag}Work done during their internships at Huawei.}
\thanks{\textsuperscript{\Letter}Corresponding authors: \{jinyibo1, fengyongxiang1, pengfei.zuo\}@huawei.com.}

\renewcommand{\shortauthors}{}

\begin{abstract}
\vspace{0.5cm}
{\centering \textnormal{\large \textbf{Abstract}} \\} 
Real-time recommender systems execute multi-stage cascades (retrieval, pre-processing, fine-grained ranking) under strict tail-latency SLOs, leaving only tens of milliseconds for ranking. Generative recommendation (GR) models can improve quality by consuming long user-behavior sequences, but in production their online sequence length is tightly capped by the ranking-stage P99 budget. We observe that the majority of GR tokens encode user behaviors that are independent of the item candidates, suggesting an opportunity to pre-infer a user-behavior prefix once and reuse it during ranking rather than recomputing it on the critical path. Realizing this idea at industrial scale is non-trivial: the prefix cache must survive across multiple pipeline stages before the final ranking instance is determined, the user population implies cache footprints far beyond a single device, and indiscriminate pre-inference would overload shared CPU/PCIe/NPU resources under high QPS.

We present \sysname, a production system that enables \textit{in-HBM relay-race inference} for GR. \sysname selectively pre-infers long-term user prefixes, keeps their per-layer KV caches resident in HBM over the request lifecycle, and ensures the subsequent ranking stage can consume them without remote fetches. \sysname combines three techniques: (1) a sequence-aware trigger that admits only \textit{at-risk} requests under a bounded cache footprint and pre-inference load, (2) an affinity-aware router that co-locates cache production and consumption by routing both the auxiliary pre-infer signal and the ranking request to the same instance, and (3) a memory-aware expander that uses server-local DRAM to capture short-term cross-request reuse while avoiding redundant reloads. We implement \sysname on Huawei Ascend NPUs and evaluate it with real queries in a production-mirror environment. Under a fixed P99 SLO, \sysname supports up to 1.5$\times$ longer sequences and improves SLO-compliant throughput by up to 3.6$\times$.

\end{abstract}

\authorsaddresses{}

\maketitle

\clearpage
\tableofcontents           
\clearpage

\input{tex/introduction}
\input{tex/background}

\input{tex/design}
\input{tex/experiment}

\input{tex/related}
\input{tex/conclusion}

\bibliographystyle{ACM-Reference-Format}
\bibliography{main}

\appendix

\end{document}

%% file: tex/introduction.tex
\section{Introduction}
\label{sec:introduction}

Modern industrial recommender systems~\cite{DBLP:journals/aim/BurkeFG11,naumov2019dlrm,DBLP:journals/corr/abs-2407-21022}
serve tens of billions of requests per day through a multi-stage cascade---retrieval, pre-processing (also known as coarse ranking), and fine-grained ranking.
Because user engagement is highly sensitive to latency, the entire pipeline must complete within a few hundred milliseconds; otherwise, timeouts translate directly into revenue loss.
The fine-grained ranking stage is the bottleneck: it typically has only tens of milliseconds at the 99th percentile (P99) to score hundreds of candidate items with a high-capacity model.

\begin{figure}[t]
    \centering
    \setlength{\abovecaptionskip}{4pt}
    \begin{subfigure}[t]{0.4\linewidth}
        \centering
        \includegraphics[width=0.85\linewidth]{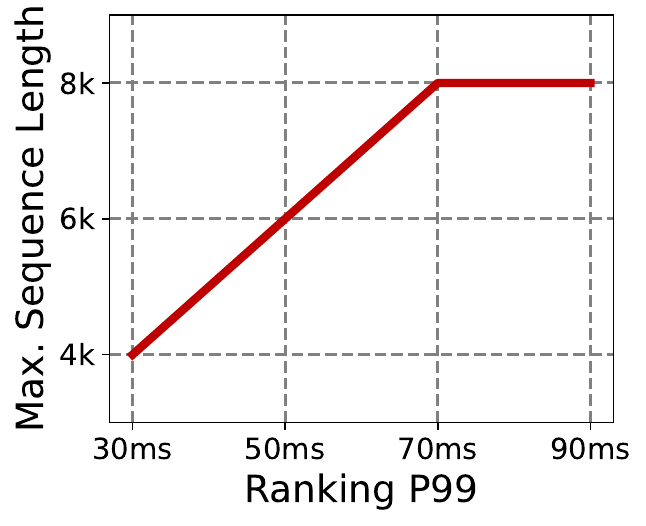}
        \caption{Restricted sequence length}
        \label{fig:1}
    \end{subfigure}
    \begin{subfigure}[t]{0.4\linewidth}
        \centering
        \includegraphics[width=0.85\linewidth]{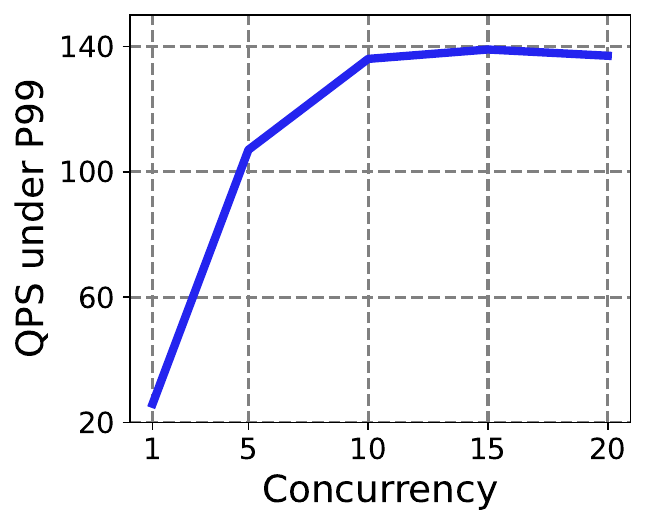}
        \caption{Restricted throughput}
        \label{fig:2}
    \end{subfigure}
    \caption{Ranking-stage P99 restricts performance.}
    \label{fig:intro_1}
\end{figure}

Generative recommendation (GR) models~\cite{DBLP:conf/nips/RajputMSKVHHT0S23,chen2024hllm,zhai2024hstu,chai2025longer,Han_2025,yan2025lum,Xu_2025,huang2025genrank,deng2025onerec, zhou2025onerecv2,liu2025onerecthink,zhang2025onetrans}
are increasingly adopted as the next generation of ranking models.
Compared with traditional deep learning recommenders (DLRMs)~\cite{DBLP:conf/kdd/WangFFW17,DBLP:conf/kdd/ZhouZSFZMYJLG18,DBLP:conf/aaai/ZhouMFPBZZG19,Huang_2019,chen2021eta,Wang_2021,zhu2025rankmixer},
GR models better capture long sequential behaviors and often exhibit favorable scaling with longer sequences and larger capacity.
Offline training and evaluation routinely use thousands of behavior tokens per user; in online serving, however, increasing sequence length (or feature dimension) quickly inflates inference latency and violates the ranking-stage P99 SLO.
Prior efforts either pursue end-to-end GR~\cite{deng2025onerec,zhou2025onerecv2,liu2025onerecthink} that unifies retrieval and ranking but may need to weaken or remove the candidate-specific cross features~\cite{Han_2025},
or deploy GR within retrieval/ranking~\cite{rajput2023tiger,Han_2025,zhang2025onetrans} and treat cross features as compensation.
In all cases, production deployments remain bounded by the same first-order barrier: ranking-stage P99 caps the usable sequence length and thus limits GR’s online scaling benefits (Figure~\ref{fig:intro_1}).

Our key observation is that GR inputs have a key structure property, i..e, most tokens encode user behaviors~\cite{DBLP:conf/ijcai/0001LGQZ0T23,zhai2024hstu,yang2025grllm} rather than candidate items.
Long-term behaviors (e.g., weeks/months of clicks and views) naturally form a prefix and evolve more slowly than short-term signals and candidate-dependent cross features injected later.
This suggests a simple opportunity: \textit{pre-infer} the long-term user prefix, cache its intermediate states (per-layer KV), and \textit{reuse} them when scoring candidate items in fine-grained ranking.
This resembles prefix caching in LLM serving~\cite{DBLP:conf/usenix/GaoHSKJDYYZ24,305212}, but the recommender setting is fundamentally harder: prefixes are user-specific (not shared across queries), and the cache must survive across multiple pipeline stages before ranking executes.

Turning the above idea into a production optimization requires solving three systems challenges.

\textbf{(1) Bridging caches across the pipeline under late binding.}
A prefix cache may be produced during retrieval, but can only be consumed during ranking after pre-processing completes.
At cache creation time, the eventual ranking instance may not yet be determined; if production and consumption land on different instances, the cache must be fetched remotely---often too expensive for a tens-of-milliseconds ranking.
Even within a server, overflowing device memory forces a flush/reload that competes with the ranking SLO.

\textbf{(2) The user population makes ``cache everything'' infeasible.}
Industrial recommenders serve billions of users~\cite{DBLP:conf/www/EksombatchaiJLL18,DBLP:conf/recsys/WangLLMWGZHBBCC24,Han_2025}.
Even modest GR backbones~\cite{zhai2024hstu,Han_2025} can generate megabytes of KV states per user when consuming thousands of tokens, yielding petabyte-scale aggregates (megabytes multiplying all users).
A naïve distributed cache pool would frequently trigger remote fetches whose latency can rival or exceed the ranking budget, undermining the entire approach.

\textbf{(3) High QPS makes unconditional pre-inference unsafe.}
At production scale, each accelerator instance may serve hundreds of queries per second~\cite{DBLP:conf/wsdm/ChenCXC21,skrlj2025dcn2}.
After feature processing and embedding lookup, each query can carry tens of megabytes of embeddings~\cite{DBLP:conf/osdi/LaiZLTWHDHPLCWR23,DBLP:conf/nsdi/YangWYWDLZZLZWD25}.
Under concurrency, host-to-device transfers and accelerator execution contend for shared CPU/PCIe/NPU resources~\cite{zuo2025cloudmatrix}.
If we pre-infer prefixes for every request, we risk overloading the very resources needed to keep ranking within P99.
Therefore, pre-inference must be selective and load-aware.

To address these challenges, we present \sysname, a production system that enables \textit{in-HBM relay-race inference} for GR.
\sysname pre-infers long-term user prefixes only when beneficial, keeps the resulting per-layer KV cache resident in device memory over the request lifecycle, and ensures the subsequent ranking stage consumes the cache locally without remote fetches.
More broadly, \sysname addresses a distinct systems problem: \textit{lifecycle caching under late-binding placement}, where cache production and consumption are separated by multiple pipeline stages and the consumer is determined only after intermediate filtering.

\sysname is built around three techniques that co-design \textit{when} to pre-infer, \textit{where} to place the cache, and \textit{how} to extend reuse safely, for the real-time recommender:

\textbf{Sequence-aware Trigger.}
During retrieval, \sysname inspects lightweight user-behavior metadata (e.g., prefix length and/or dimension) and estimates whether the request is \textit{at risk} of violating the ranking-stage P99 under full inference.
Only at-risk requests are admitted for prefix pre-inference, bounding the overhead and preventing it from becoming a new bottleneck.

\textbf{Affinity-aware Router.}
For admitted requests, \sysname enforces a routing contract: the subsequent ranking request must arrive at the same instance that produced the cache.
The design leverages the fact that a request lifecycle lasts only a few hundred milliseconds, allowing HBM to act as a sliding-window cache that retains per-user prefixes long enough for consumption (e.g., to enable the relay-race), while eliminating cross-server fetches on the ranking critical path.

\textbf{Memory-aware Expander.}
\sysname further exploits server-local DRAM to capture short-term reuse across repeated requests from the same user (e.g., rapid refresh) without introducing remote traffic.
HBM guarantees cache availability within a single lifecycle, while DRAM acts as a controlled compensation tier that extends reuse beyond the HBM window at bounded H2D cost.

We implement \sysname on Huawei Ascend NPUs and evaluate it with real queries under production constraints.
Under a fixed ranking-stage P99 SLO (also the entire recommender SLO), \sysname supports up to 1.5$\times$ longer input sequences while improving the system throughput by up to 3.6$\times$.

In summary, this paper makes three contributions:
\begin{itemize}
    \item \textit{Problem and insight.} We identify ranking-stage P99 as a first-order barrier to deploying long-sequence GR online and observe that most GR computation lies in an item-independent user-behavior prefix.
    \item \textit{System design.} We present \sysname, a production system that realizes in-HBM relay-race inference as lifecycle caching under late-binding placement via the sequence-aware trigger, the affinity-aware routing, and the DRAM-backed expander.
    \item \textit{Production implementation and evaluation.} We implement \sysname on Ascend NPUs and demonstrate up to 1.5$\times$ longer supported sequences and up to 3.6$\times$ higher SLO-compliant throughput under fixed P99 SLOs.
\end{itemize}

%% file: tex/background.tex
\begin{figure}[!t]
    \centering
    \includegraphics[width=0.8\linewidth]{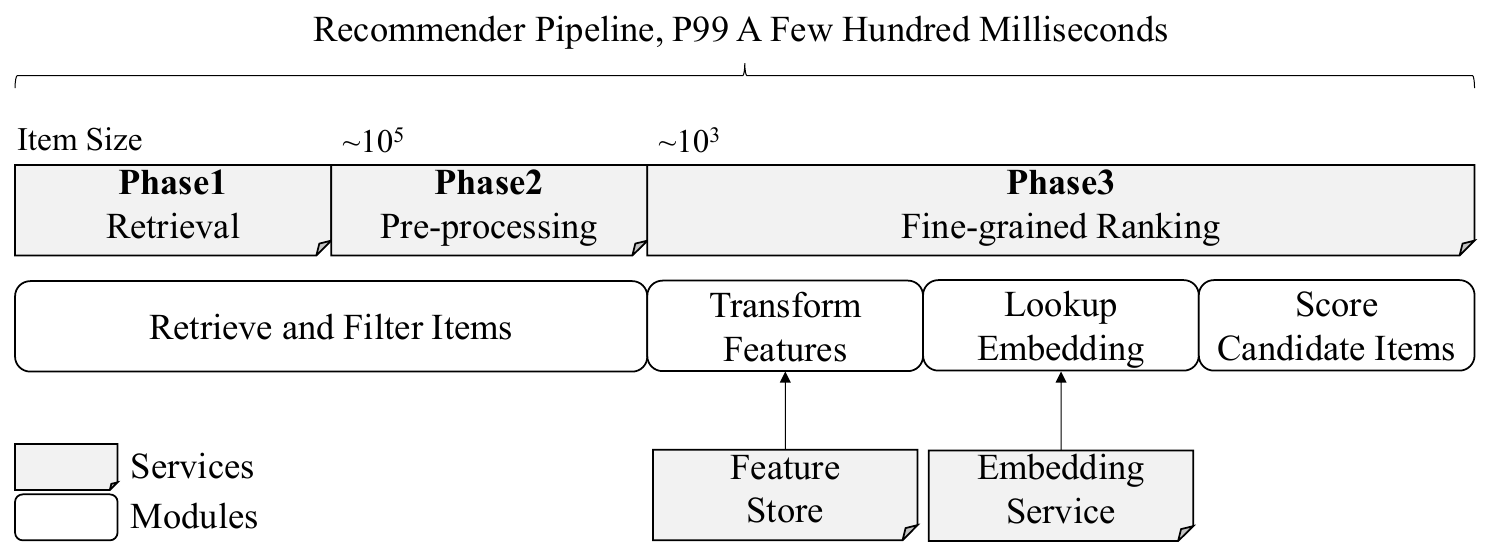}
    \caption{Recommender with multiple phases.}
    \label{fig:intro_2}
\end{figure}

\section{Background and Motivation}

Real-time recommender systems actually execute a multi-stage cascade---retrieval, pre-processing, and fine-grained ranking---under strict tail-latency SLOs. Meanwhile, generative recommendation (GR) models benefit from long user-behavior sequences, but production deployments cap online sequence length (and also the feature dimension) to meet the ranking-stage P99 budget. This section first reviews the real-time pipeline (\S\ref{sec:background:pipeline}) and the structure of GR inference (\S\ref{sec:background:gr}). We then motivate an opportunity to pre-infer user-behavior prefixes (\S\ref{sec:background:opportunity}) and explain why turning that opportunity into an industrial-scale optimization is non-trivial (\S\ref{sec:background:challenges}).

\subsection{Real-time Recommender Pipeline}
\label{sec:background:pipeline}

\paragraph{Latency budgets and tail amplification.}
Industrial recommenders process massive traffic volumes and must keep end-to-end latency within a few hundred milliseconds at P99 to avoid timeout-induced failures and revenue loss~\cite{on_device_reranking,xi2025seral}. Crucially, the budget is partitioned across pipeline stages, leaving only tens of milliseconds for fine-grained ranking. Because ranking runs at high concurrency, production deployments also avoid operating near peak utilization: saturating CPU, PCIe, or accelerators amplifies queuing delay and worsens tails, making P99 compliance brittle.

\paragraph{Three-stage cascade.}
A typical pipeline comprises: (i) \textit{retrieval}, which selects a large candidate pool; (ii) \textit{pre-processing/coarse ranking}, which performs feature transformation and prunes candidates; and (iii) \textit{fine-grained ranking}, which scores the remaining candidates with heavyweight models~\cite{DBLP:journals/aim/BurkeFG11,naumov2019dlrm,DBLP:journals/corr/abs-2407-21022}. Figure~\ref{fig:intro_2} illustrates this cascade. Candidate set size shrinks stage by stage, but ranking remains latency-critical because it must score hundreds of items within a rigid P99 window while competing for shared CPU/PCIe/accelerator resources with other pipeline work.

\subsection{Generative Recommendation Under Tight P99}
\label{sec:background:gr}

\paragraph{Why GR is attractive.}
GR models improve recommendation quality by modeling long sequential user behaviors and exhibiting favorable scaling with longer sequences and larger model capacity~\cite{DBLP:conf/nips/RajputMSKVHHT0S23,chen2024hllm,zhai2024hstu,chai2025longer,Han_2025,yan2025lum,Xu_2025,huang2025genrank,deng2025onerec,zhou2025onerecv2,liu2025onerecthink,zhang2025onetrans}. In offline training and evaluation, GR models can consume thousands of behavior tokens per user, learning rich dependencies across a user’s behavior trajectory and the contents presented over time.

GR can be deployed in two common modes. In a \textit{discriminative} mode, the model outputs scores/logits for candidate ranking, effectively learning decision boundaries between items~\cite{chai2025longer,Han_2025,chen2024hllm}. In a \textit{generative} mode, the model produces targets token by token or generates representations that are consumed by downstream towers for retrieval or ranking~\cite{rajput2023tiger,zhai2024hstu,zhang2025onetrans}. While recent end-to-end GR proposals aim to unify retrieval and ranking~\cite{deng2025onerec,zhou2025onerecv2,liu2025onerecthink}, in practice, candidate-dependent cross features remain an important source of quality gain~\cite{Han_2025}. Many deployments therefore still rely on them (explicitly or implicitly) even when adopting GR backbones.

\begin{figure}[!t]
    \centering
    \includegraphics[width={0.7\linewidth}]{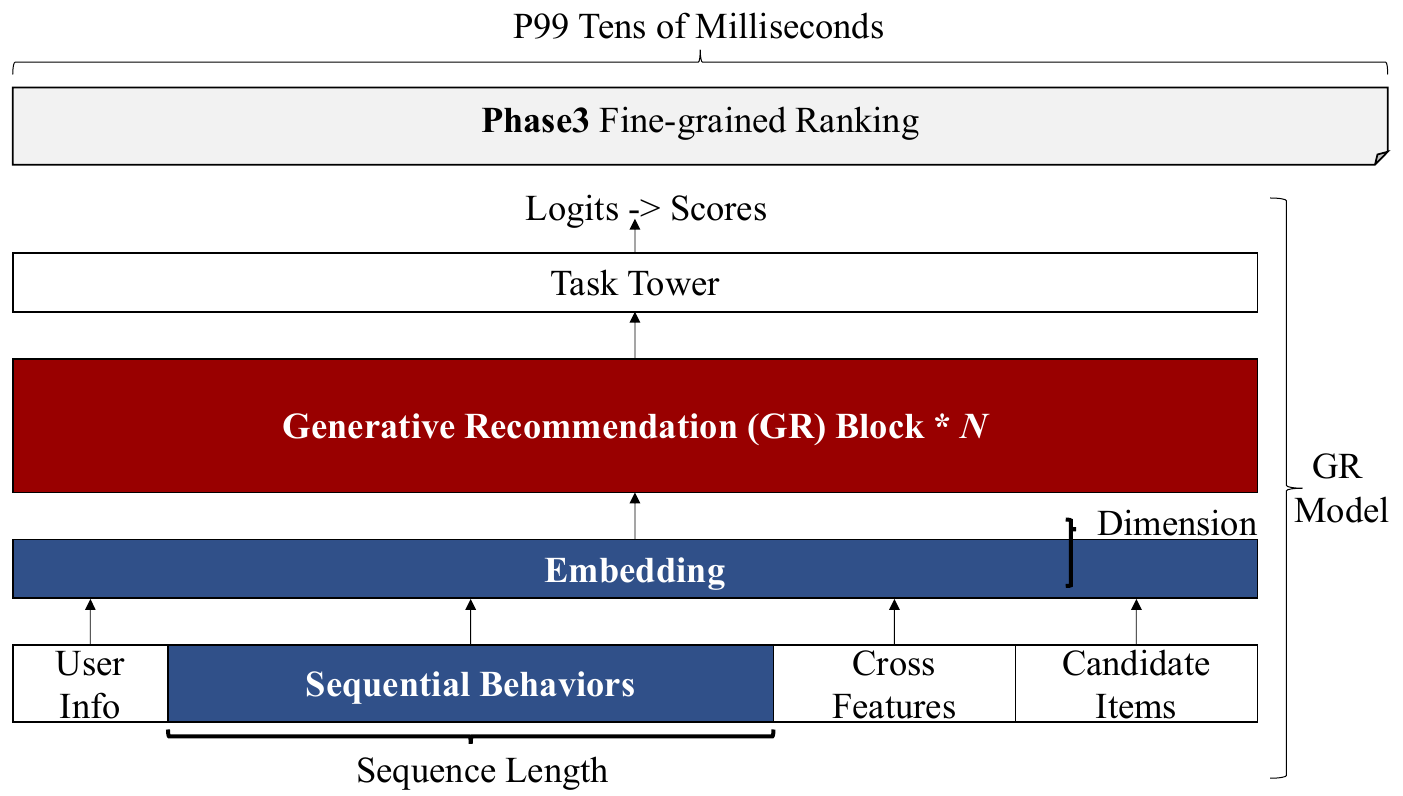}
    \caption{Limited sequences (length or dimension).}
    \label{fig:back_1}
\end{figure}

\paragraph{Why GR is constrained online.}
Online fine-grained ranking operates under a \textit{tens-of-milliseconds} P99 budget. Increasing GR sequence length or feature dimension directly increases inference cost and makes tail latency hard to sustain. As a result, production systems cap the online sequence length, creating a gap between offline training (thousands of behaviors) and online inference (typically far fewer)~\cite{zhai2024hstu,Han_2025,deng2025onerec,chen2024hllm,zhang2025onetrans}. Existing responses largely \textit{shrink} compute to fit the fixed budget---for example, compressing tokens or representations~\cite{chai2025longer} (Figure~\ref{fig:back_1}). For industrial deployments, however, realizing GR’s scaling benefits requires the complementary capability: \textit{raising the online sequence-length ceiling} while still meeting the same tail-latency SLO.

\subsection{Opportunity: Pre-infer User-behavior Prefix}
\label{sec:background:opportunity}

\paragraph{Long-term behaviors form a large, stable prefix.}
GR inputs interleave user-side context and item-side targets. In practice, user behaviors are time-ordered; long-term behaviors (months/years to days ago) appear early and constitute a large prefix~\cite{Han_2025,zhang2025onetrans,deng2025onerec}. The remaining tokens typically include short-term behaviors (hours/days) and cross features, followed by the candidate items to be scored. Importantly, long-term behaviors dominate the token count (often thousands) and evolve slowly, whereas short-term signals and embeddings are refreshed frequently and are tied to tight update pipelines (e.g., hourly refresh).
This paper therefore focuses on pre-inferring \textit{only} the long-term prefix: it captures the dominant compute while avoiding strong coupling with fast model/embedding refresh and preserving production robustness~\cite{10.1145/3725783.3764389,DBLP:conf/osdi/LaiZLTWHDHPLCWR23,DBLP:conf/eurosys/XieLLWGRS22,Wei_2022}.

\begin{figure}[!t]
    \centering
    \includegraphics[width={0.7\linewidth}]{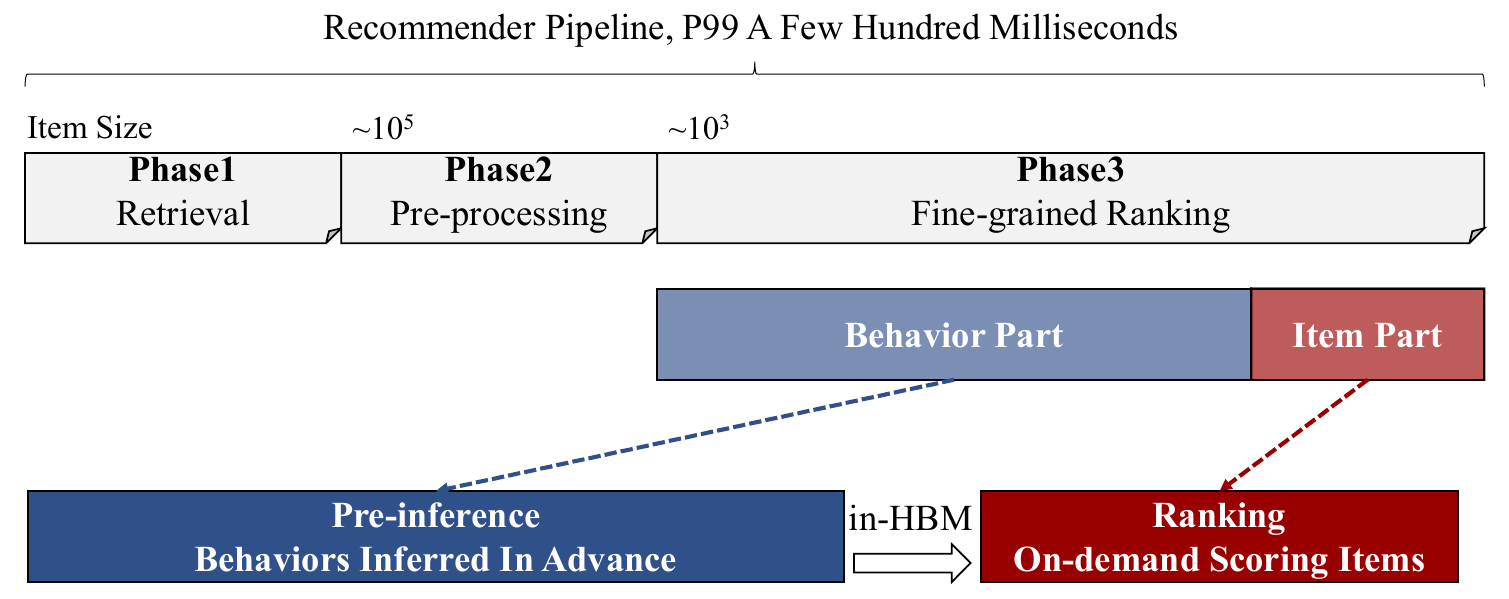}
    \caption{Behaviors inferred in advance and reused at ranking.}
    \label{fig:back_2}
\end{figure}

\paragraph{Prefix reuse: analogous to LLMs, but with different reuse semantics.}
Prefix caching is widely used in LLM inference to reuse computation across requests~\cite{gim2024promptcache,ye2024chunkattention,DBLP:conf/usenix/GaoHSKJDYYZ24,305212}. In LLMs, prefixes are often \textit{shared} (common prompts/templates), so reuse is cross-request and cross-user. In GR, prefixes are predominantly \textit{user-specific}: two users rarely share the same behavior history. Thus, the most reliable reuse opportunity is \textit{within-user} reuse across stages of the same request lifecycle (and, opportunistically, across repeated requests from the same user), not broad cross-user caching.

\paragraph{Pre-infer before ranking.}
The long-term prefix is available at the beginning of retrieval and is largely decoupled from candidate filtering. This enables a ``race-ahead'' execution: compute the prefix during retrieval, cache its intermediate states, and reuse them when ranking later scores candidate items (Figure~\ref{fig:back_2}). If the expensive long-prefix computation is removed from the ranking critical path, ranking latency becomes much less sensitive to total sequence length; it primarily depends on the incremental tokens (short-term/cross features) and the item batch.
We formalize this idea as
\begin{small}
\begin{equation}
\psi \leftarrow f([\mathcal{U}, \mathcal{S}_{l}, \emptyset, \emptyset], \emptyset),\;\;\;
\left|f([\mathcal{U}, \mathcal{S}_{l}, \widetilde{\mathcal{S}_{l}}, \mathcal{I}], \emptyset) - 
f([\emptyset, \emptyset, \widetilde{\mathcal{S}_{l}}, \mathcal{I}], \psi)\right| \le \varepsilon,
\nonumber
\end{equation}
\end{small}
where $f$ is the GR ranking model (a generative backbone plus downstream task towers).
The input sequence contains user info $\mathcal{U}$, long-term behaviors $\mathcal{S}_{l}$, non-long-term tokens $\widetilde{\mathcal{S}_{l}}$ (short-term behaviors $\mathcal{S}_{s}$ and cross features $\mathcal{C}$; either may be $\emptyset$), and candidate items $\mathcal{I}$.
$\psi$ is the cached per-layer KV state produced by pre-inferring the long-term prefix.
The bound $\varepsilon$ indicates that using $\psi$ preserves ranking scores (up to a small deviation) relative to full inference.

\subsection{Challenges: Making Prefix Caching Work at Industrial Scale}
\label{sec:background:challenges}

Pre-inferring long-term prefixes is appealing, but making it effective \textit{online} requires solving three systems challenges.

\paragraph{(1) Cross-stage cache survivability under late-binding placement.}
In LLM serving, the cache produced by prefill is consumed immediately by decode, and the system can hand it off at a clear phase boundary~\cite{patel2024splitwise,zhong2024distserve}. In a recommender pipeline, a GR prefix cache may finish during retrieval or pre-processing but is only consumed when fine-grained ranking begins. At production time, the final ranking instance is often chosen \textit{after} pre-processing, so cache producer and consumer are separated by both time and routing. If the consumer differs from the producer, a remote fetch is required; if the cache is spilled to host memory, D2H/H2D adds latency and contention. Either path can consume a material fraction of a tens-of-milliseconds ranking budget. An effective solution must therefore ensure the cache survives across stages and is consumed locally.

\paragraph{(2) Footprint at billion-user scale makes ``cache everything'' impossible.}
Accelerator HBM provides only tens of GBs. Even for relatively small GR models, a per-user prefix cache can be MB-scale at a few-thousand-token input; at billion-user scale, the aggregate is inevitably TB--PB. This eliminates any design that assumes persistent per-user residency in device memory and makes naïve distributed KV pools problematic: remote fetch latency can match or exceed ranking-stage P99. Therefore, an online solution must avoid remote cache fetches on the ranking critical path and treat caching as a \textit{lifecycle} mechanism rather than a global persistent store.

\paragraph{(3) Operational safety under high QPS and shared-resource contention.}
Production traffic reaches tens of billions of requests per day, and each instance commonly sustains hundreds of QPS. Under such concurrency, CPU feature processing, H2D transfers (tens of MBs of embeddings per request), and accelerator execution compete for shared resources. Unconditionally pre-inferring every request is infeasible: it increases PCIe/NPU pressure and can \textit{worsen} ranking tail latency. Instead, the system must (i) admit only \textit{at-risk} requests that would otherwise violate ranking-stage P99, and (ii) bound the admitted pre-inference rate and live-cache footprint in a service-specific way, trading available slack in retrieval against added load near ranking.

\paragraph{Implication.}
An effective design must selectively pre-infer long-term prefixes, enforce affinity between cache production and consumption to eliminate remote fetch, and tightly control the additional compute and memory pressure under high QPS.
This motivates the \sysname design in \S\ref{sec:design}.

%% file: tex/design.tex
\section{The \sysname Design}
\label{sec:design}

\subsection{System Overview}
\label{overview}

\sysname addresses the \textit{industrial barrier} for long-sequence GR serving: ranking must satisfy a tens-of-milliseconds P99 budget, yet most GR compute lies in a user-behavior prefix that is available early and is independent of candidate items. \sysname introduces a \textit{relay-race} side path that computes and reuses prefix states across the recommender pipeline (retrieval $\rightarrow$ pre-processing $\rightarrow$ ranking), without adding remote cache fetches to the ranking critical path.

\paragraph{Abstraction: lifecycle caching under late-binding placement.}
We model prefix reuse as \textit{lifecycle caching}. For a request associated with user key $u$, the system may produce a prefix cache $\psi(u)$ early and consume it later:
\begin{itemize}
    \item \textbf{Cache object.} $\psi(u)$ is the \textit{per-layer KV cache} of the GR backbone computed on the long-term behavior prefix of $u$. Importantly, $\psi(u)$ is a deterministic function of the prefix tokens and model weights, and does not depend on candidate items.
    \item \textbf{Lifecycle window.} Let $T_{\text{life}}(u)$ denote the time from when the auxiliary \texttt{pre-infer} is issued to when the corresponding ranking consumes $\psi(u)$. In production, $T_{\text{life}}$ is bounded by the pipeline tail (retrieval + pre-processing + ranking), and is typically a few hundred milliseconds.
    \item \textbf{Late-binding placement.} The eventual ranking instance is chosen after pre-processing. Thus, at the time $\psi(u)$ is produced, the consumer location is not yet fixed unless the system enforces a placement contract.
\end{itemize}
The system's objective is to make $\psi(u)$ \textit{available} throughout $T_{\text{life}}(u)$ and \textit{locally consumable} at ranking time, under high QPS and bounded device memory.

\begin{figure}[!t]
    \centering
    \includegraphics[width=0.8\linewidth]{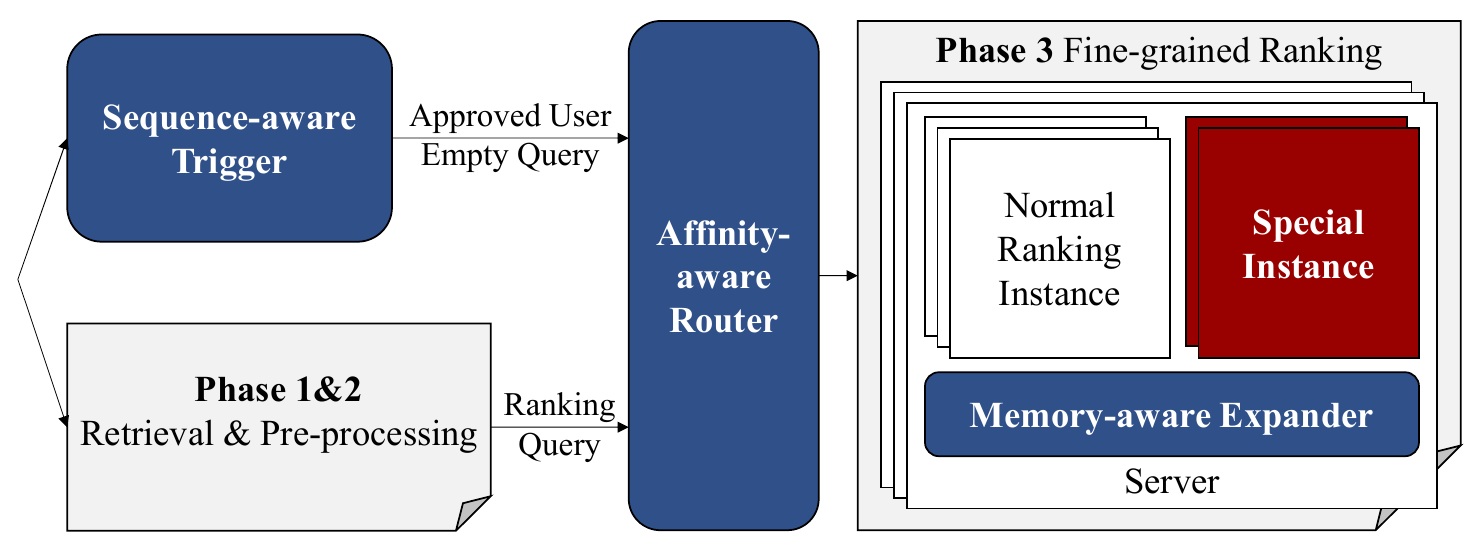}
    \caption{System overview of \sysname.}
    \label{fig:design_1}
\end{figure}

\paragraph{System invariants.}
\sysname is designed to enforce two invariants for every \textit{admitted} long-sequence request:
\begin{itemize}
    \item \textbf{(I1) No remote fetch on ranking critical path.} Ranking either consumes $\psi(u)$ locally (HBM or server-local DRAM) or safely falls back to baseline inference; it never blocks on cross-server cache fetching.
    \item \textbf{(I2) Survivability under bounded footprint and load.} The system caps (a) the live-cache footprint in HBM and (b) the admitted pre-inference rate so that the caches survive for at least $T_{\text{life}}$ while preserving P99 SLOs.
\end{itemize}

\paragraph{Decomposition: admission, placement, and local capacity extension.}
As shown in Figure~\ref{fig:design_1}, \sysname decomposes lifecycle caching into three responsibilities:
\begin{enumerate}
    \item \textbf{Admission (\textit{sequence-aware trigger}).} Decide \textit{which} requests should create $\psi(u)$ (only ``at-risk'' long-sequence requests) and \textit{how many} can be admitted so their caches survive $T_{\text{life}}$ under HBM and rate budgets (enforcing I2).
    \item \textbf{Placement (\textit{affinity-aware router}).} Convert late-binding routing into an early-binding placement contract: ensure the auxiliary \texttt{pre-infer} and later ranking request for the same $u$ land on the \textit{same} special instance, so ranking can reuse $\psi(u)$ without remote fetch (enforcing I1).
    \item \textbf{Local capacity extension (\textit{memory-aware expander}).} Opportunistically extend reuse across repeated requests from the same user using \textit{server-local} DRAM, while bounding DRAM$\rightarrow$HBM reload overhead and avoiding redundant reloads under concurrency (preserving I1 and stabilizing tail latency).
\end{enumerate}

\paragraph{End-to-end flow.}
If the trigger admits user $u$, it issues a response-free \texttt{pre-infer} signal to a special ranking service, causing the selected instance to compute $\psi(u)$ and keep it resident in HBM for $T_{\text{life}}(u)$. The router then guarantees that the subsequent ranking request for $u$ is routed to the same instance, allowing ranking to reuse $\psi(u)$ locally. After consumption, the expander may spill $\psi(u)$ to server-local DRAM to accelerate short-term repeated requests (e.g., rapid refresh), with bounded reload concurrency and correctness-preserving ordering control.

In summary, the design goal of this paper is to provide a \textit{systems contract} for GR prefix reuse: admitted caches survive across pipeline stages and are consumed locally at ranking time under strict constraints.
By separating \textit{what to compute} (admission) from \textit{where to consume} (placement) and \textit{how to extend reuse locally} (DRAM tier), \sysname turns prefix caching into a controlled mechanism that preserves correctness while avoiding remote fetches and tail-latency blowups.

\begin{figure}[!t]
    \centering
    \includegraphics[width=0.75\linewidth]{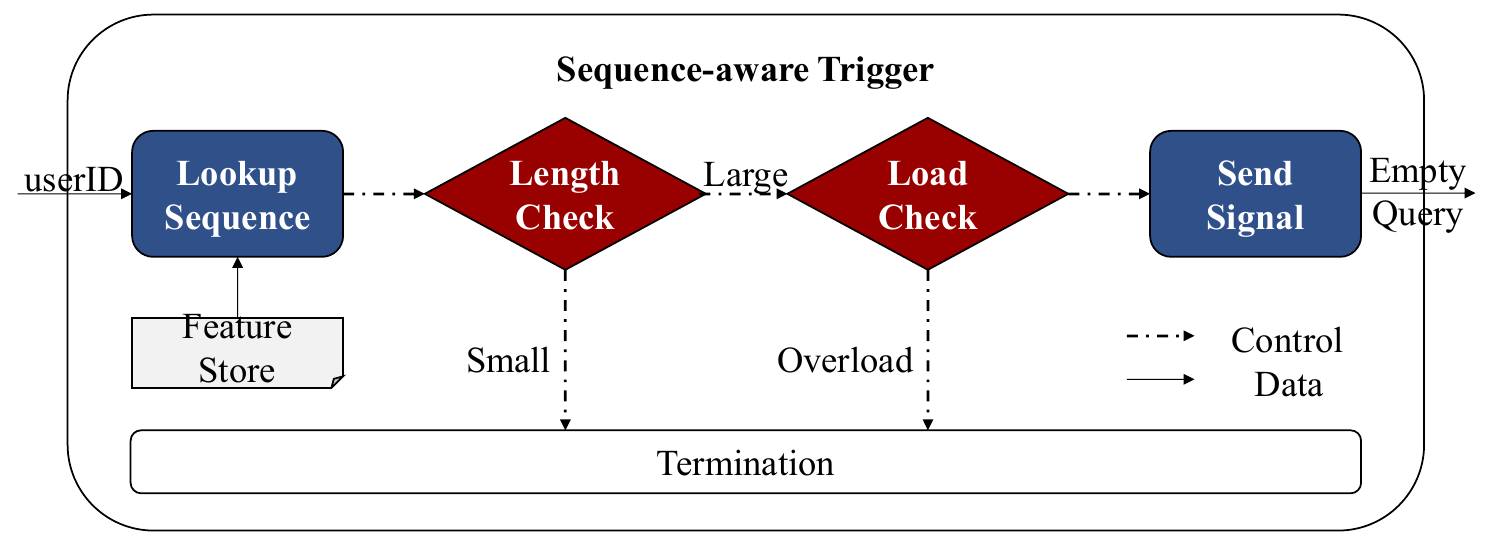}
    \caption{Workflow of sequence-aware trigger.}
    \label{fig:design_2}
\end{figure}

\subsection{Sequence-aware Trigger}
\label{trigger}

\paragraph{Design goal.}
Admit only \textit{at-risk} long-sequence requests for prefix pre-inference and ensure their caches $\psi$ can \textit{survive} until ranking consumption, subject to bounded HBM footprint and bounded additional pre-inference load.

\paragraph{Side-path risk test using metadata.}
Figure~\ref{fig:design_2} illustrates the trigger workflow. The trigger runs in parallel with retrieval and only fetches lightweight behavior metadata (e.g., prefix length and/or feature dimension). This avoids transferring full behavior sequences and keeps the trigger fast. If metadata indicates that full ranking inference is unlikely to violate the ranking-stage P99, the trigger terminates immediately and introduces \textit{zero} additional work. Only requests deemed \textit{at risk} (e.g., long prefixes or high-dimensional features) are candidates for admission.

\paragraph{Admission control via lifecycle-window survivability.}
Pre-inference is beneficial only if $\psi$ remains available when ranking arrives. Let $Q_{\text{admit}}$ be the admitted \texttt{pre-infer} rate (queries/s) for a given special instance. Over a lifecycle window of length $T_{\text{life}}$, the number of \textit{simultaneously live} caches per special instance is upper-bounded by
\begin{equation}
L \;\triangleq\; Q_{\text{admit}} \cdot T_{\text{life}}.
\label{eq:livecache}
\end{equation}
Using $kv_{p99}$ as the P99 footprint of $\psi$ (per admitted user) and reserving a fraction $r_1$ of HBM for live caches, survivability requires the following inequality per special instance:
\begin{equation}
L \cdot kv_{p99} \;\le\; r_1 \cdot HBM.
\label{eq:hbm_budget}
\end{equation}

\paragraph{Bounding pre-inference load.}
Let $Q_m$ be the sustainable \texttt{pre-infer} throughput per model slot (queries/s) on a special instance and $M$ the number of concurrent model slots. The trigger caps admission by
\begin{equation}
Q_{\text{admit}} \;\le\; Q_m \cdot M,\qquad
Q_{\text{max}} \;\le\; (Q_m \cdot M)\cdot (r_2 \cdot N),
\label{eq:qps_budget}
\end{equation}
where $N$ is the total number of ranking instances and $r_2N$ are designated as special. The first inequality prevents per-instance overload; the second bounds system-wide admitted long-sequence traffic by the aggregate capacity of the special-instance pool (i.e., $r_{2}$ is a control variable).

\begin{figure}[!t]
    \centering
    \includegraphics[width=0.75\linewidth]{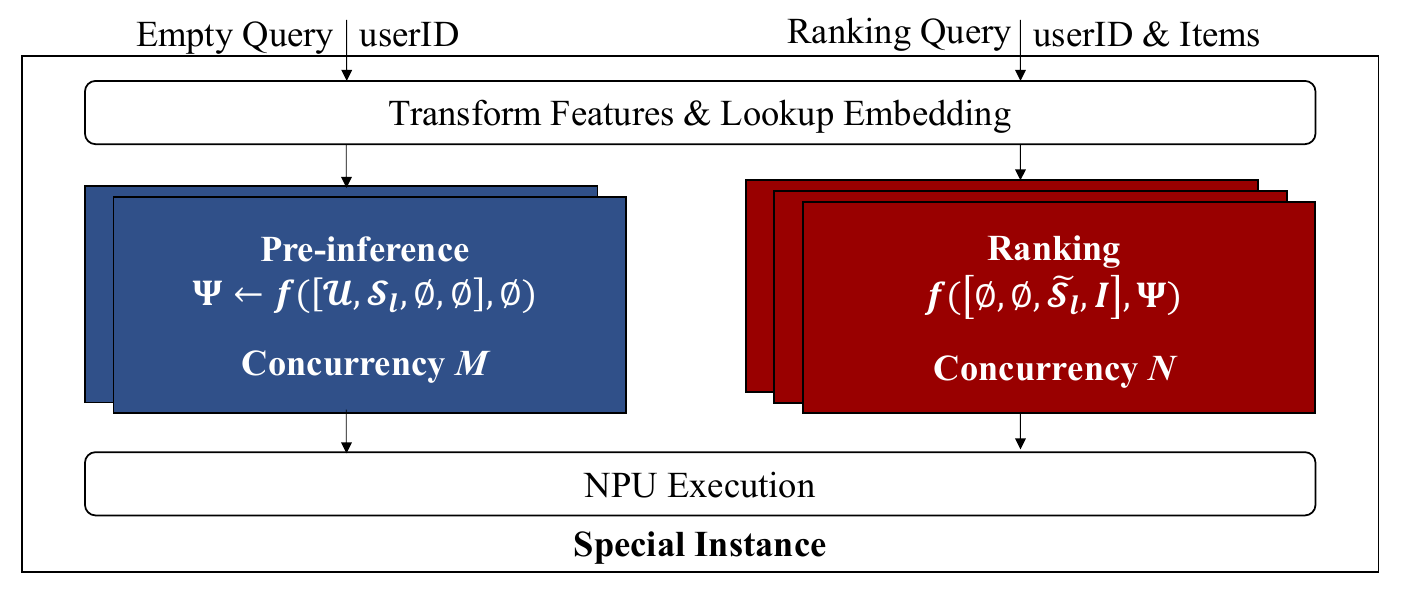}
    \caption{Inference within ``special'' instance.}
    \label{fig:design_3}
\end{figure}

\paragraph{Example (sanity check).}
If pre-inference takes 35\,ms, then $Q_m \approx \lfloor 1000/35 \rfloor \approx 30$ queries/s per model slot. With $M{=}5$, $kv_{p99}{\approx}0.1$\,GB, $HBM{=}32$\,GB, and $r_1{=}0.5$, Eq.~\ref{eq:hbm_budget} allows up to $L \le 16/0.1 = 160$ live caches per special instance, corresponding to $Q_{\text{admit}} \le 160/T_{\text{life}}$ via Eq.~\ref{eq:livecache}. Meanwhile Eq.~\ref{eq:qps_budget} caps compute: $Q_{\text{admit}} \le 30 \cdot 5 = 150$ QPS per special instance. With $N{=}100$ and $r_2{=}0.1$, the pool supports $Q_{\text{max}}\le 1500$ QPS of admitted long-sequence traffic.

\paragraph{Auxiliary \texttt{pre-infer} request as a control signal.}
For an admitted request, the trigger sends a response-free \texttt{pre-infer} signal carrying the user-keyed consistency hash:
\begin{verbatim}
  header { consistency-hash-key: userID, ... }
  body   { user_id: userID, stage: pre-infer }
\end{verbatim}
This signal asks the special instance to fetch the user’s long-term behaviors, run pre-inference, and populate $\psi$ in HBM. Because it is issued during retrieval, it contains no candidate items (obey the same format of ranking queries).

\paragraph{Concurrency inside a special instance.}
A special instance processes a mix of auxiliary \texttt{pre-infer} requests and ranking requests for different users (Figure~\ref{fig:design_3}). Increasing $M$ improves overlap but also increases contention on compute and memory-transfer engines; thus, $M$ is a tunable knob jointly configured with $r_2$ and admission thresholds. Upon request arrival, the instance checks \texttt{stage}: if \texttt{pre-infer}, it computes and stores $\psi$; otherwise, it looks up $\psi$ by user ID and performs ranking using the cached prefix when present.

\paragraph{Why it works.}
The trigger converts an unbounded optimization into a bounded one: only requests that would otherwise violate ranking-stage P99 are admitted, and Eqs.~\ref{eq:livecache}--\ref{eq:qps_budget} ensure admitted caches survive for $T_{\text{life}}$ without overloading HBM or pre-inference capacity (invariant I2).

\begin{figure}[!t]
    \centering
    \includegraphics[width=0.75\linewidth]{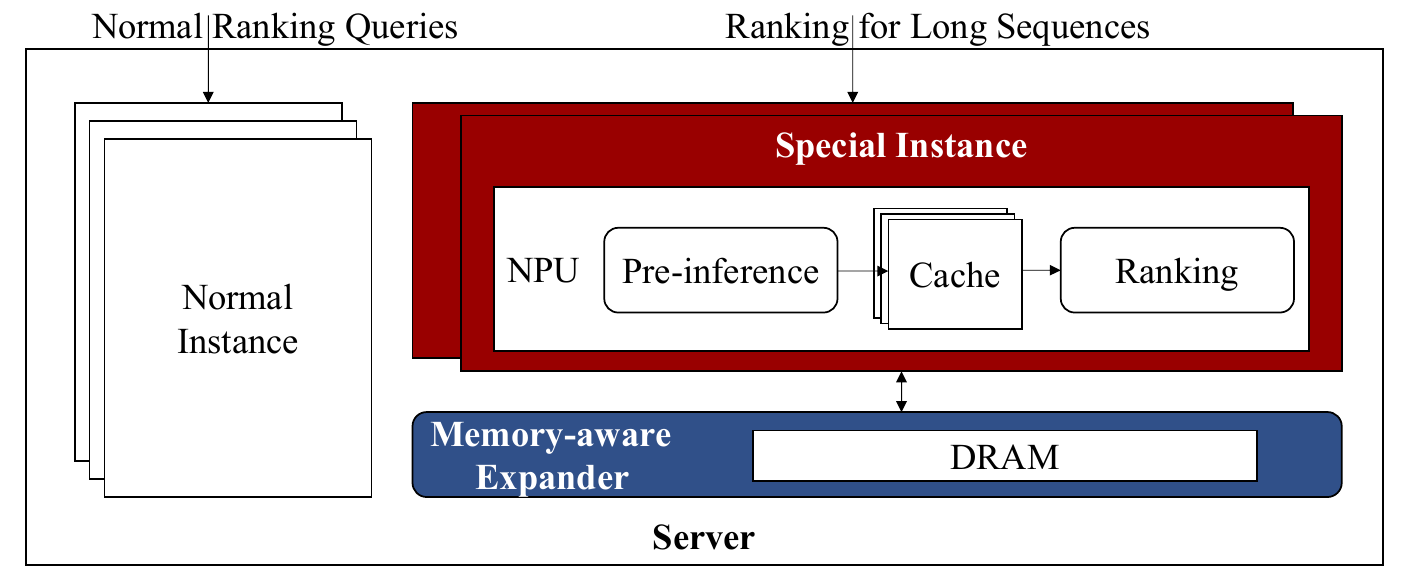}
    \caption{Instance deployment.}
    \label{fig:design_4}
\end{figure}

\subsection{Affinity-aware Router}
\label{router}

\paragraph{Design goal.}
Guarantee \textit{placement}: for every admitted user, the auxiliary \texttt{pre-infer} request and the subsequent ranking request are routed to the \textit{same} special instance, so ranking can reuse $\psi$ without cross-server fetching (invariant I1).

\paragraph{Special-instance pool and interference control.}
As illustrated in Figure~\ref{fig:design_4}, \sysname deploys both normal and special ranking instances across servers, distinguished by the service names. To bound interference on shared server resources, we cap the number of special instances per server (typically one or two). This matters because pre-inference stresses (i) CPU (behavior processing to construct model inputs) and (ii) PCIe/H2D bandwidth (transferring embeddings from host DRAM to device HBM). Even when a server is partitioned into multiple instances, PCIe remains a shared bottleneck; thus, bounding special-instance density per server limits worst-case contention.

\begin{figure}[!t]
    \centering
    \includegraphics[width=0.75\linewidth]{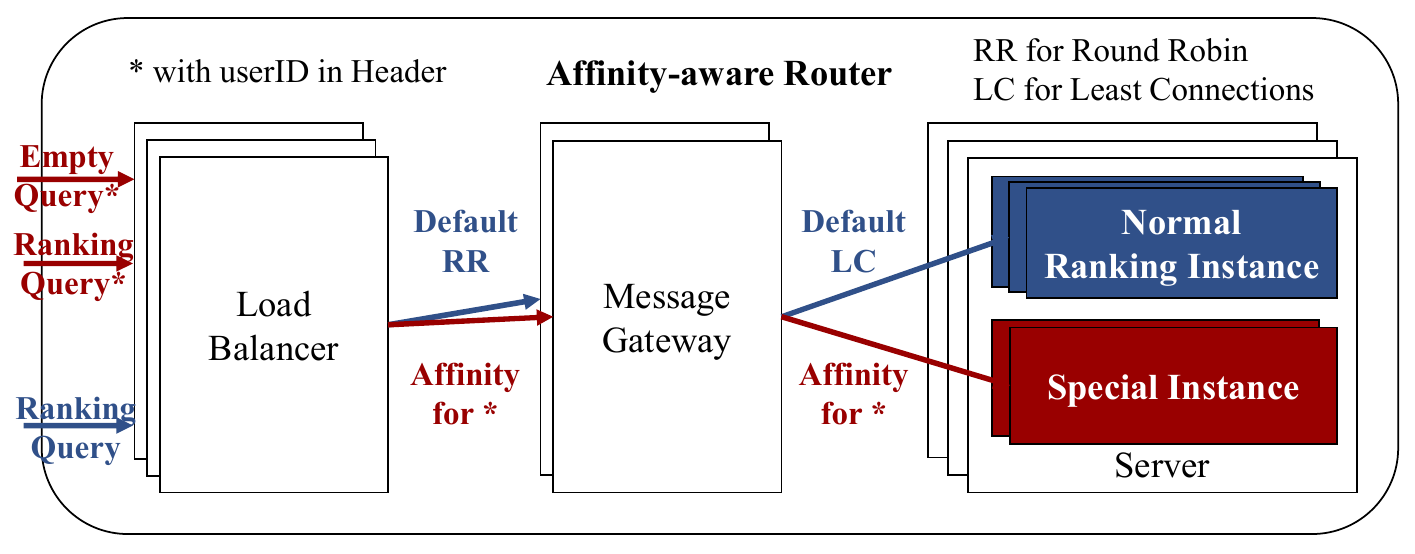}
    \caption{Routing via load balancer and gateway.}
    \label{fig:design_5}
\end{figure}

\paragraph{Balancing independent requests vs.\ coupling related requests.}
Long-sequence traffic is distributed across special instances by the existing load balancer and gateway (Figure~\ref{fig:design_5}). Standard policies (round-robin, least-connections) balance \textit{independent} requests. However, \sysname introduces \textit{coupled} requests for the same user: an auxiliary \texttt{pre-infer} and a later ranking request. To reuse $\psi$, these two must rendezvous at the same cache-holding instance.

\paragraph{Affinity routing to exact the cache-holding instance.}
For long-sequence users, the ranking request carries a \texttt{consistency-hash-key} derived from the ID in its header:
\begin{verbatim}
  header { consistency-hash-key: userID, ... }
  body   { user_id: userID, items: ... }
\end{verbatim}
The body includes both the user ID and the filtered items required for ranking. During pre-processing, the system determines whether a request should be served by normal or special instances: short-sequence requests follow the normal service, while long-sequence requests are sent to the special service. For special instances, the user ID also serves as the lookup key for the HBM-resident prefix cache; including it in both header and body improves robustness (e.g., for cache-miss fallback paths that require additional processing).

Both the auxiliary \texttt{pre-infer} request and the long-sequence ranking request carry the same key in the header \texttt{consistency-hash-key}. The load balancer and gateway apply \textit{consistent hashing} on this key when selecting the gateway and the final instance, respectively, ensuring that the two related requests are routed to the same recipient. If affinity is disrupted (e.g., deployment churn or network changes) and the ranking request lands on a different instance, that instance simply falls back to inference without cache, preserving correctness at the cost of losing the optimization. For normal ranking requests that do not include the key, the system continues to use standard load-balancing policies (e.g., round-robin and least-connections).

\begin{figure}[!t]
    \centering
    \includegraphics[width=0.75\linewidth]{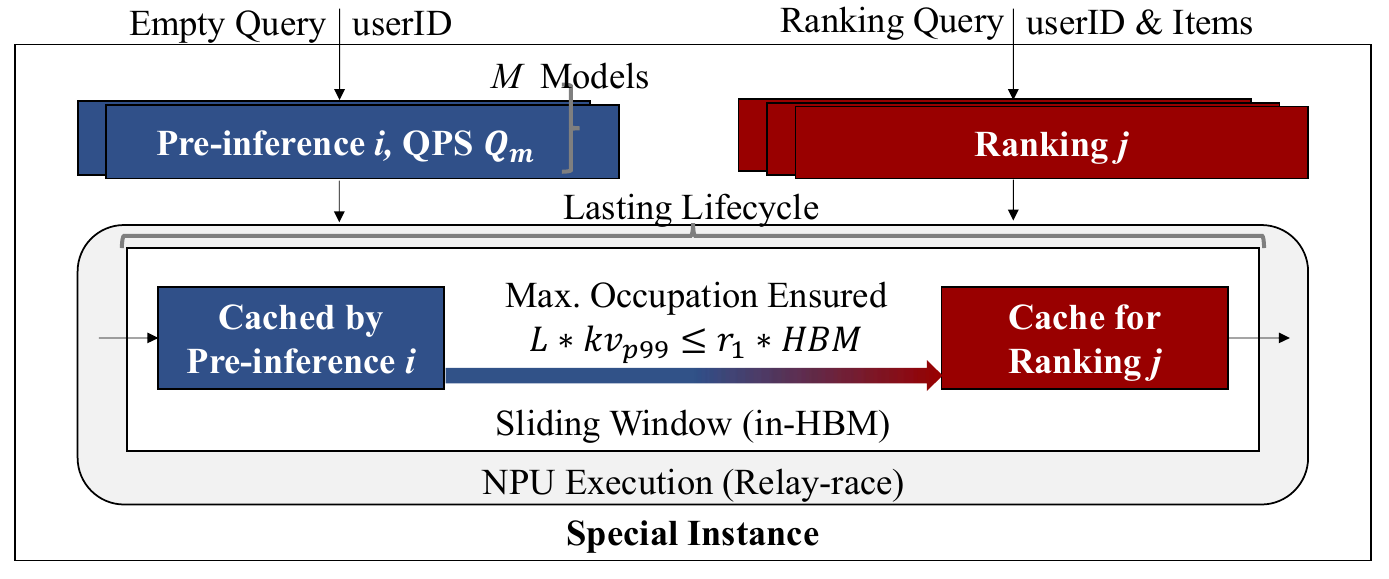}
    \caption{HBM forms sliding window.}
    \label{fig:design_6}
\end{figure}

\paragraph{HBM as a lifecycle window.}
As enforced by the trigger, the total HBM footprint consumed by prefix caches is bounded. Consequently, HBM functions as a \textit{sliding-window cache} (Figure~\ref{fig:design_6}): per-user prefix caches are inserted by pre-inference, consumed by ranking, and then evicted as new admitted users arrive. Although HBM cannot hold caches for all users, admission control ensures it can hold the caches that must survive for a single recommendation lifecycle---typically a few hundred milliseconds. Since the cache lifetime is short, HBM capacity is sufficient to cover the required window even with model concurrency.

Finally, the ratio of special instances to all ranking instances determines how long-sequence load is distributed: increasing the ratio reduces per-instance pressure (CPU/PCIe/HBM) and enlarges the effective window, while decreasing it concentrates long-sequence traffic onto fewer instances. In practice, we set this ratio to be small (often $<10\%$), roughly matching the fraction of long-sequence users, and rely on autoscaling to keep per-instance load stable as traffic grows.

\paragraph{Why it works.}
Consistent hashing turns late-binding placement into a stable \textit{affinity contract}: producer and consumer rendezvous at the same instance without coordination, eliminating remote fetch on the ranking critical path (invariant I1). If affinity is disrupted (e.g., churn), the system safely falls back to baseline inference, preserving correctness.

\subsection{Memory-aware Expander}
\label{expander}

\paragraph{Design goal.}
Extend reuse of $\psi$ across repeated requests from the same user uses server-local DRAM, while bounding DRAM$\rightarrow$HBM reload overhead and preventing redundant reloads under concurrent and out-of-order arrivals.

\paragraph{DRAM as a server-local reuse tier.}
HBM residency is sufficient to bridge caches \textit{within} a single recommendation lifecycle: once admitted, the prefix cache remains available until the corresponding ranking consumes it. However, the same user may trigger multiple recommendation trials in a short period (e.g., rapid refresh), and HBM alone cannot retain caches across these trials while simultaneously serving many other users. In contrast, host DRAM provides hundreds of GBs of capacity per server. \sysname therefore uses DRAM as a \textit{server-local} reuse tier: after a cache is consumed, it can be spilled to DRAM and reloaded later to accelerate subsequent requests, without incurring cross-server fetches.

Upon receiving a ranking request at a special instance, \sysname performs a two-level lookup using the user ID as the key: it first checks HBM, and on a miss it queries the local expander for a DRAM hit. If the cache exists in DRAM, the instance reloads it into HBM (incurring H2D) and proceeds to ranking. To prevent the expander from becoming a new bottleneck, DRAM-to-HBM reloads are rate-limited and scheduled with bounded concurrency.

Concurrent requests for the same user can otherwise trigger repeated DRAM-to-HBM reloads and repeated cache state transitions (spill/reload). The expander enforces \textit{per-user serialization}: it maintains a per-user queue (or lock) and ensures that at most one cache-affecting action is in-flight per user at any time. Subsequent requests for the same user wait for the first operation to complete, then reuse the cache in HBM.

\paragraph{Handling out-of-order arrivals.}
In practice, auxiliary pre-inference and ranking for the same user may arrive out of order (Figure~\ref{fig:design_7}). For example, if behavior processing is slow, the \texttt{pre-infer} request may be delayed while one or more ranking requests (with different candidate sets) arrive earlier. A naïve design would cause each ranking request to reload the cache from DRAM independently, leading to redundant H2D transfers and increased tail latency.

To handle out-of-order arrivals and eliminate redundant reloads, \sysname inserts a lightweight \textit{pseudo pre-inference} step in front of each ranking request. Conceptually, the per-user queue becomes:
\textit{pseudo-pre-infer $\rightarrow$ rank $\rightarrow$ pseudo-pre-infer $\rightarrow$ rank $\rightarrow \dots$}
The pseudo step performs the same cache checks as real pre-inference: it first probes HBM, then DRAM on a miss. If the cache is in DRAM, only the \textit{first} pseudo step triggers a DRAM-to-HBM reload; all subsequent pseudo steps hit in HBM and proceed directly to ranking. This guarantees that, even when the real \texttt{pre-infer} request arrives late (e.g., after several ranking requests), the system performs at most one reload per user and avoids redundant transfers while preserving correctness.

\paragraph{Why it works.}
Server-local DRAM increases effective cache capacity without violating the no-remote-fetch invariant (I1). The per-user single-flight serialization and idempotent pseudo-pre-inference ensure at-most-once DRAM$\rightarrow$HBM reload per user per burst, eliminating redundant transfers and stabilizing tail latency under concurrency.

%% file: tex/experiment.tex
\section{Performance Evaluation}

\label{sec:eval}

We evaluate \sysname in a production-mirror recommender environment to answer three questions:
\begin{itemize}
    \item \textbf{Q1 (Effectiveness).} Does in-HBM relay-race inference improve tail-latency compliant capacity, i.e., the \textit{maximum supported sequence length} and the \textit{SLO-compliant throughput}?
    \item \textbf{Q2 (Scaling).} As sequence length grows, does \sysname degrade gracefully and continue to provide non-trivial throughput under the same P99 constraints?
    \item \textbf{Q3 (Extensibility).} Does \sysname remain effective when extending GR models (e.g., larger embedding dimensions and deeper backbones), and when varying deployment knobs (e.g., candidate set size and concurrency)?
\end{itemize}

\subsection{Prototype and Experimental Setup}

\label{sec:eval:setup}

\paragraph{Prototype in a production-mirror environment.}
We deploy \sysname as a set of production-compatible services in a mirror environment and evaluate it with real queries. The recommender pipeline follows the standard cascade: retrieval $\rightarrow$ pre-processing (coarse ranking) $\rightarrow$ fine-grained ranking. Each phase typically consumes tens of milliseconds; the pipeline-level tail-latency SLO is P99 $\le 135$\,ms, and fine-grained ranking is the tightest stage (e.g., $\sim$50\,ms budget at P99). Fine-grained ranking consists of (i) feature processing, (ii) embedding lookup via external embedding service, and (iii) GR inference on NPUs.

\paragraph{Deployment and routing substrate.}
Requests are forwarded through the same load balancers and message gateways as in production, which we extend to support affinity-aware routing via a user-keyed \texttt{consistency-hash-key}. The network is tenant-isolated, while the load balancers and gateways are shared among multiple services. We deploy multiple instances of both load balancers and gateways to match production fan-out and contention patterns.

\paragraph{Ranking instances and ``special'' instances.}
Fine-grained ranking uses two service names: \textit{normal} instances and \textit{special} instances (for over-long sequences, e.g., length larger than a configured threshold such as 4K). Each ranking instance (normal or special) is mapped to one Ascend 910C NPU with tens-of-GB HBM. Instances process multiple requests concurrently by combining CPU-side parallelism (feature/sequence processing) and NPU-side model concurrency (multiple model slots). The sequence-aware trigger controls the admitted pre-inference load and thus the effective QPS delivered to special instances.

\paragraph{Models and workloads.}
We evaluate multiple GR architectures used in practice and their variants (e.g., HSTU~\cite{zhai2024hstu} and its revised version, and our internal GR blocks). Across experiments, sequence lengths range from 1K to tens of thousands of tokens; embedding dimensions range from 128 to 1024; and the generative backbone contains several to tens of layers. These settings stress the ranking critical path due to high FLOP and memory costs under long sequences. The distribution of the user-behavior length shows that most users have short histories and fewer than 6\% have long sequences exceeding 2K tokens. Even caching only these long-sequence users would require tens of terabytes of storage, far beyond the capacity of any single device. 

\paragraph{Cache semantics.}
The cached object $\psi$ is the per-layer KV cache of the generative backbone computed from long-term behavior prefix (e.g., $\sim$32\,MB for a 2K-token request in HSTU with 8 layers, 256-dimensional embeddings, and fp32 format). \sysname keeps $\psi$ in HBM over the request lifecycle window, and optionally spills it to server-local DRAM for short-term cross-request reuse.

\paragraph{Metrics and measurement protocol.}
We report:
(1) \textit{Maximum supported sequence length}, defined as the largest sequence length that meets the pipeline SLO (P99 $\le 135$\,ms) with a \textbf{success rate} $\ge 99.9\%$ across all evaluated queries;
(2) \textit{SLO-compliant throughput} (QPS) per special instance and at the system level, measured under the same P99 constraints; and
(3) \textit{Component latencies} (P99) for \texttt{pre} (pre-inference), \texttt{load} (DRAM$\rightarrow$HBM cache loading), and \texttt{rank} (ranking with cache).
We distinguish \textit{QPS} (completed queries per second) from \textit{concurrency} (simultaneously in-flight queries), since the latter is the key driver of resource contention and tail latency.

\begin{figure*}[t!]
    \begin{subfigure}[t]{0.4\textwidth}
        \includegraphics[width=0.9\textwidth]{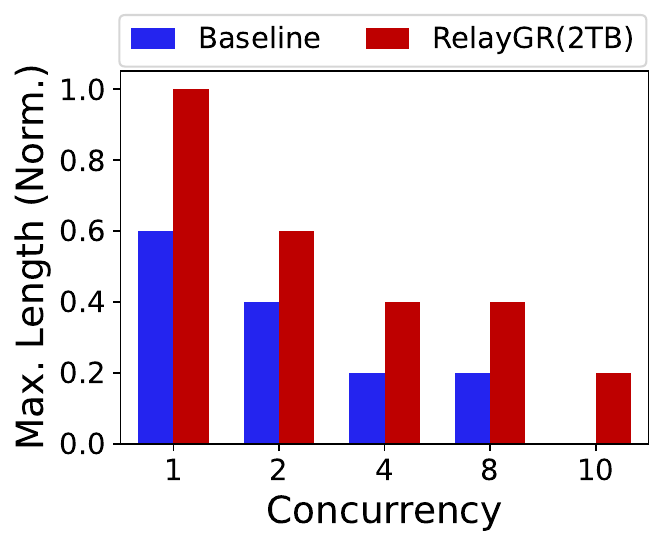}
        \caption{Maximum sequence length supported.} 
        \label{fig:exp1_1}
    \end{subfigure}
    \begin{subfigure}[t]{0.4\textwidth}
        \includegraphics[width=0.9\textwidth]{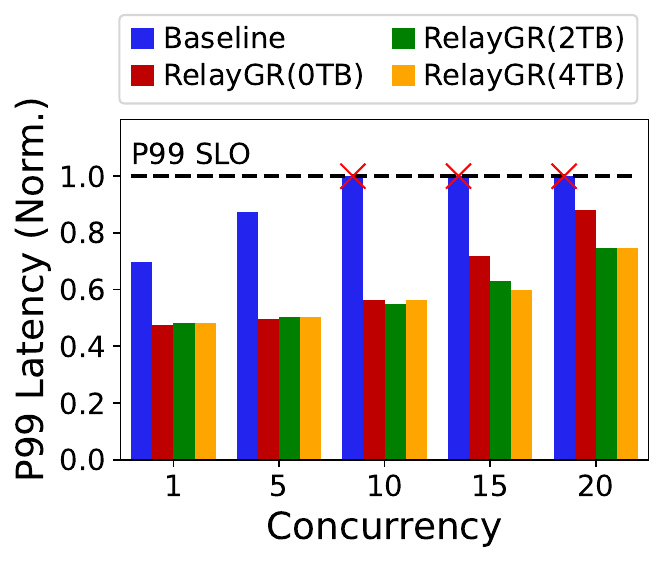}
        \caption{P99s under various concurrency.}
        \label{fig:exp1_2}
    \end{subfigure}
    
    \begin{subfigure}[t]{0.4\textwidth}
        \includegraphics[width=0.9\textwidth]{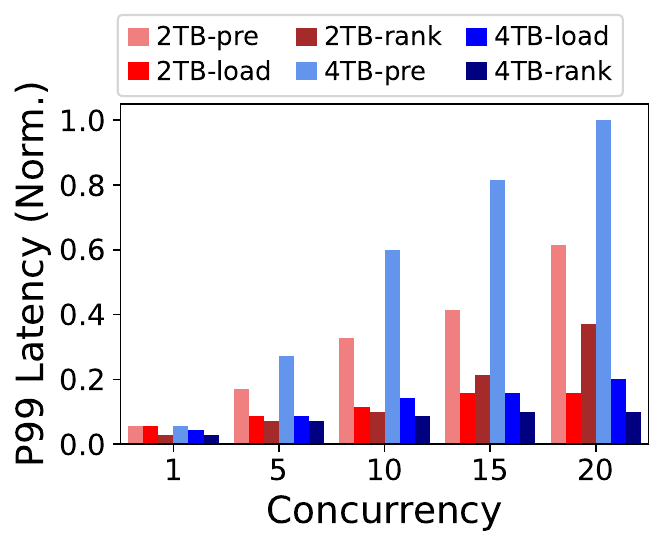}
        \caption{Details under various concurrency.} 
        \label{fig:exp1_3}
    \end{subfigure}
    \begin{subfigure}[t]{0.4\textwidth}
    \includegraphics[width=0.9\textwidth]{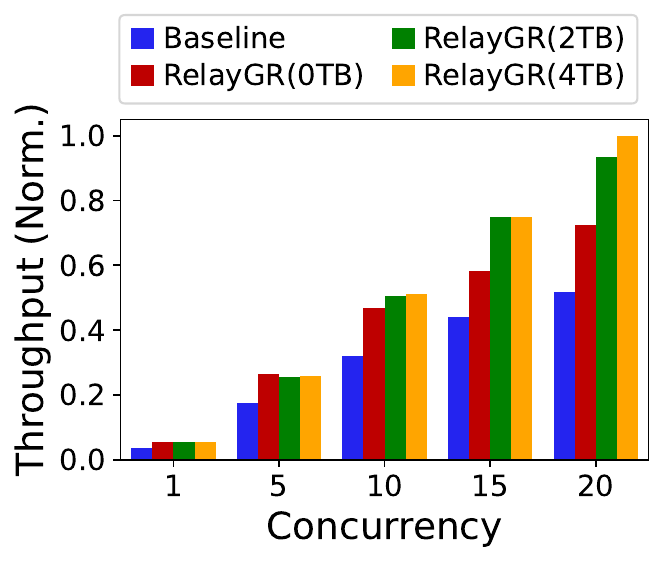}
        \caption{Throughput under various concurrency.} 
        \label{fig:exp1_4}
    \end{subfigure}
    \caption{Evaluation on the effectiveness of \sysname.}
\end{figure*}

\paragraph{Baselines and variants.}
\textit{Baseline} is the production configuration without relay-race: fine-grained ranking performs full GR inference inline under the ranking-stage P99.
\textit{\sysname} denotes in-HBM relay-race inference with no DRAM reuse (0\% DRAM hit).
\textit{\sysname+$x$\%} denotes \sysname augmented with server-local DRAM reuse where $x\%$ is the measured DRAM hit rate (controlled by available DRAM budget for spilling, typically 500GB and maximum 4TB).

\subsection{Q1: Effectiveness of \sysname}
\label{sec:eval:rq1}

\paragraph{Maximum supported sequence length.}
Figure~\ref{fig:exp1_1} shows that \sysname increases the maximum sequence length that satisfies the P99 SLO compared to the baseline. Adding server-local DRAM reuse further extends the maximum supported length by avoiding repeated pre-inference across bursts of requests from the same users. In our setup, using all available DRAM ($\sim$500GB) yields a DRAM hit rate of about 10\% and increases the maximum supported sequence length by up to \textbf{1.5$\times$} over the baseline. (We discuss higher hit rates enabled by additional tiers, e.g., SSD, as an extension point; the core results in this paper focus on HBM/DRAM, i.e., 2TB and 4TB yielding $\sim$50\% and $\sim$100\% hit rates, respectively.)

\paragraph{Tail latency under concurrency.}
Figure~\ref{fig:exp1_2} reports end-to-end P99 under different concurrency levels at fixed sequence length. The baseline quickly violates the pipeline SLO as concurrency increases because full GR inference remains on the ranking critical path. \sysname shifts the expensive long-prefix computation to the relay-race path, enabling roughly \textbf{2$\times$} more concurrent in-flight requests without exceeding the P99 SLO in our experiments (also confirmed in our production environment). Increasing DRAM hit rate further reduces compute on the special instances, improving tail latency headroom at the same concurrency.

\paragraph{Where the savings come from.}
Figure~\ref{fig:exp1_3} breaks down P99 latency into \texttt{pre}, \texttt{load}, and \texttt{rank}. Pre-inference cost grows rapidly with sequence length because it processes the long-term prefix; this is precisely why removing it from the ranking critical path is essential. In contrast, \texttt{load} and \texttt{rank} grow much more slowly: cache loading is linear in cache size, and ranking-on-cache only processes the incremental tokens and candidate items. Consequently, even as concurrency increases, \sysname keeps the ranking tail latency below the baseline by confining the expensive component to the relay-race path. The \texttt{rank} further decreases using half precision (i.e., fp16 format) or the NPU type with higher computing power.

\begin{figure}[t]
    \begin{subfigure}[h]{0.4\textwidth}
        \centering
        \setlength{\abovecaptionskip}{4pt}
        \includegraphics[width=\textwidth]{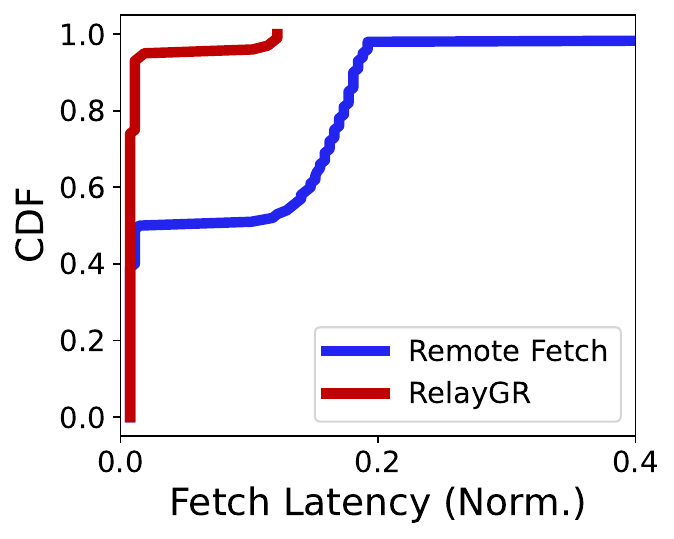}
    \end{subfigure}
    \caption{Comparison of local (\sysname) and remote fetch.}
    \label{fig:exp_extra1}
\end{figure}

\paragraph{SLO-compliant throughput.}
Figure~\ref{fig:exp1_4} shows that \sysname improves throughput (QPS) under the P99 constraint. With DRAM reuse, more requests skip pre-inference and directly reuse cached $\psi$, yielding further gains. Using all available DRAM improves throughput by up to \textbf{3.6$\times$} over the baseline in our setup. These results indicate that relay-race inference is a practical path to making long-sequence GR cost-effective in production while maintaining strict P99 SLOs.


\begin{figure*}[t!]
    \begin{subfigure}[t]{0.4\textwidth}
        \includegraphics[width=0.9\textwidth]{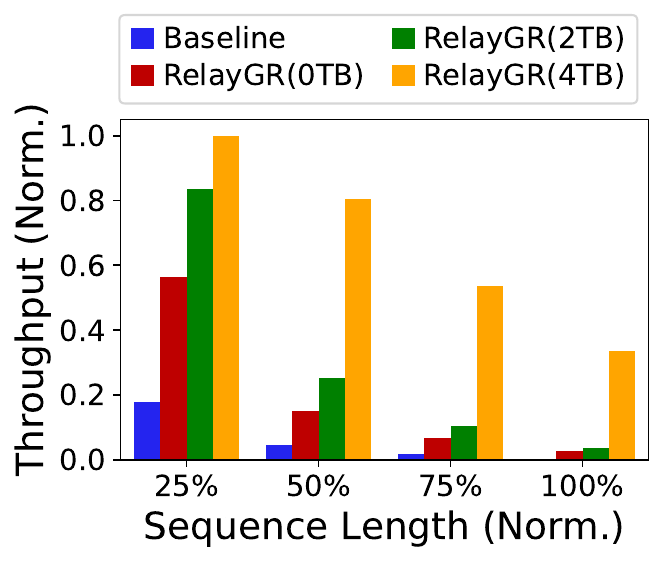}
        \caption{Throughput under various sequences.} 
        \label{fig:exp2_1}
    \end{subfigure}
    \begin{subfigure}[t]{0.4\textwidth}
        \includegraphics[width=0.9\textwidth]{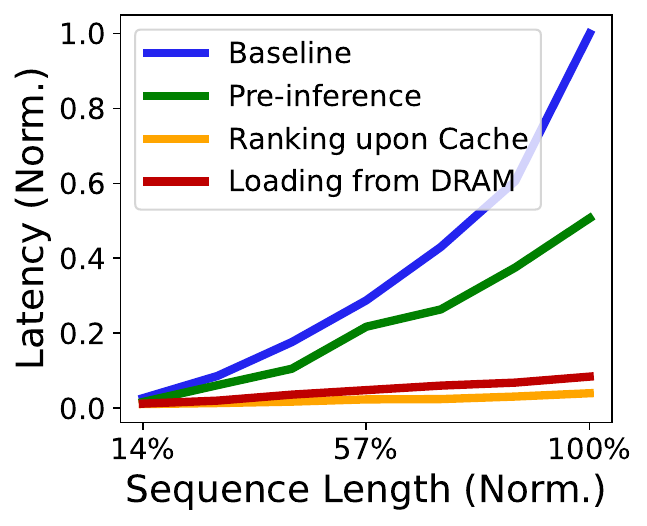}
        \caption{Latency under various sequences.}
        \label{fig:exp2_2}
    \end{subfigure}
    
    \begin{subfigure}[t]{0.4\textwidth}
        \includegraphics[width=0.9\textwidth]{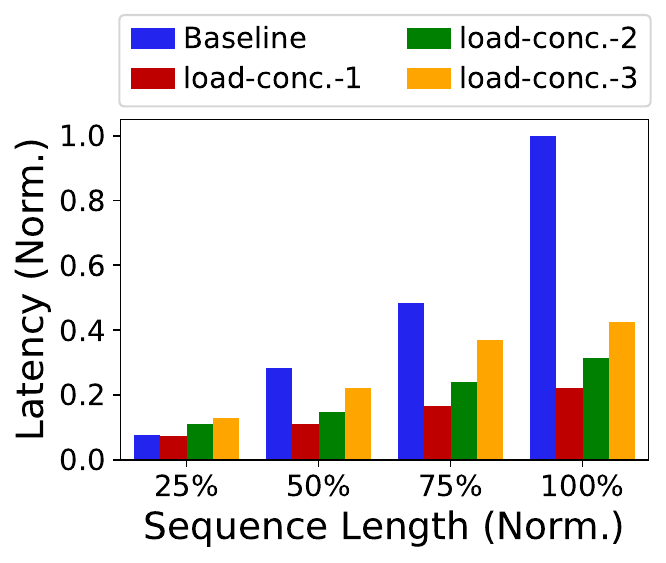}
        \caption{Loading under various sequences.} 
        \label{fig:exp2_3}
    \end{subfigure}
    \begin{subfigure}[t]{0.4\textwidth}
        \includegraphics[width=0.9\textwidth]{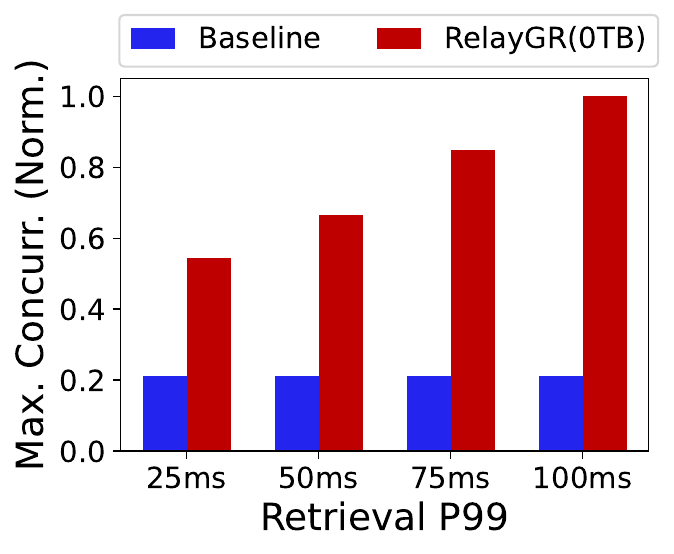}
        \caption{Maximum concurrency.} 
        \label{fig:exp2_4}
    \end{subfigure}
    \caption{Evaluation on \sysname for scaled sequences.}
\end{figure*}

\begin{figure*}[t!]
    \begin{subfigure}[t]{0.4\textwidth}
        \includegraphics[width=0.9\textwidth]{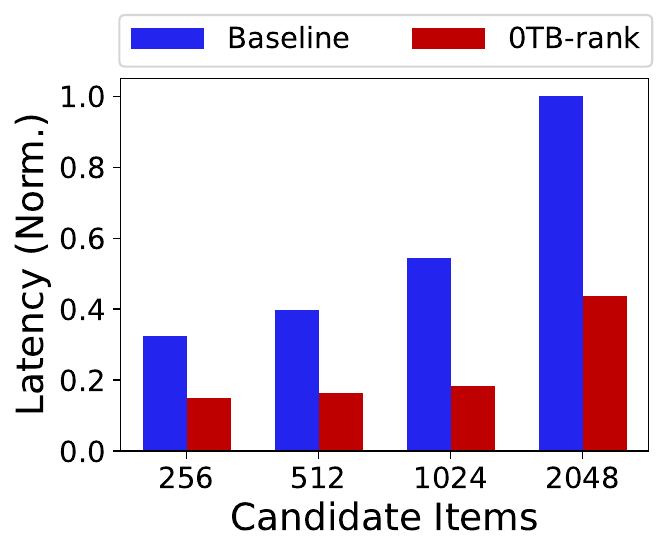}
        \caption{Latency under various items.} 
        \label{fig:exp3_1}
    \end{subfigure}
    \begin{subfigure}[t]{0.4\textwidth}
        \includegraphics[width=0.9\textwidth]{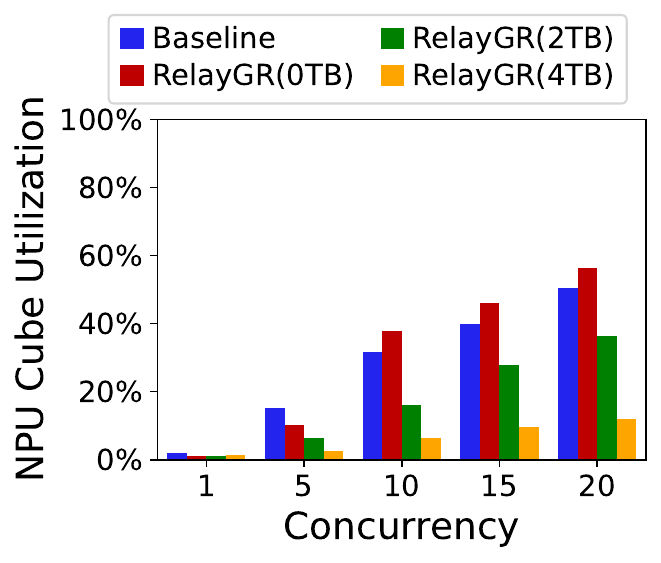}
        \caption{Cube util. under various concurrency.}
        \label{fig:exp3_2}
    \end{subfigure}
    
    \begin{subfigure}[t]{0.4\textwidth}
        \includegraphics[width=0.9\textwidth]{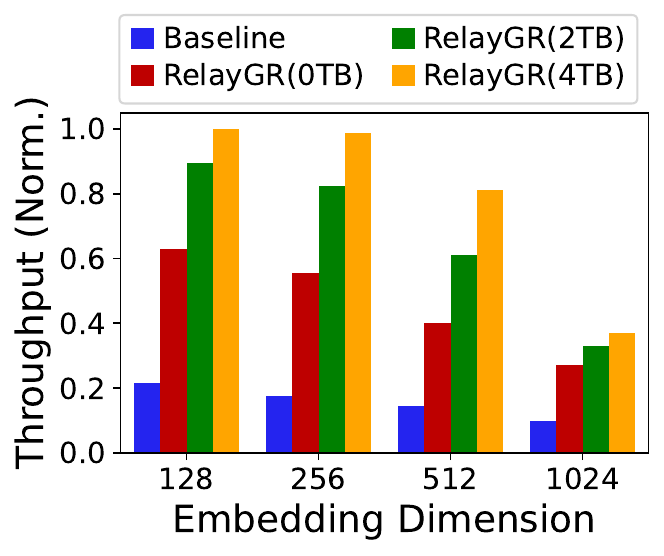}
        \caption{Throughput under various dimensions.} 
        \label{fig:exp3_3}
    \end{subfigure}
    \begin{subfigure}[t]{0.4\textwidth}
        \includegraphics[width=0.9\textwidth]{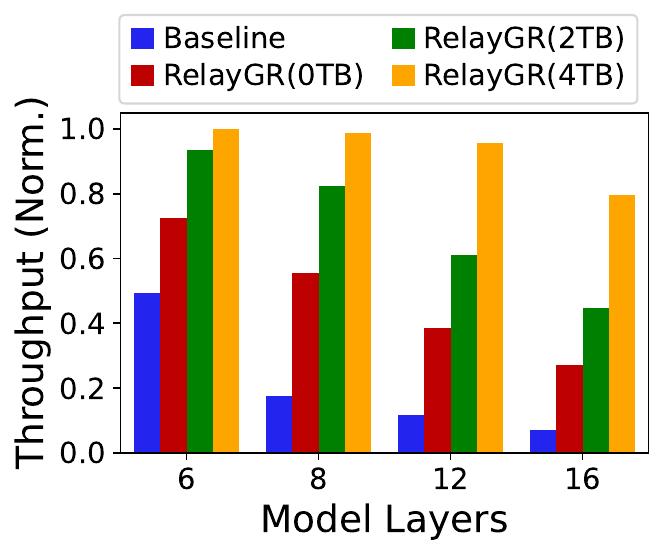}
        \caption{Throughput under various model layers.} 
        \label{fig:exp3_4}
    \end{subfigure}
    \caption{Evaluation on the extension of \sysname.}
\end{figure*}

\paragraph{Affinity is necessary.}
If prefix caches are placed in a distributed pool without affinity (i.e., without \sysname), ranking may require remote cache fetches across servers. As shown in Figure~\ref{fig:exp_extra1}, remote fetch latency (blue line) can be hundreds of times higher than local-cache access (red line, \sysname) and can easily exceed the request lifecycle window.. These results motivate \sysname’s key invariant: \textit{no remote fetch on the ranking critical path}.

\subsection{Q2: \sysname for Scaled Sequences}
\label{sec:eval:rq2}

\paragraph{Throughput degrades gracefully with longer sequences.}
Figure~\ref{fig:exp2_1} shows that throughput decreases as sequence length grows for all approaches, but \sysname degrades much more slowly than the baseline. The baseline throughput collapses to only a few QPS beyond $\sim$6K tokens under the same SLO (far from the throughput demands), while \sysname maintains \textit{tens of QPS} even with 0\% DRAM hit (pure in-HBM relay-race) beyond 6K tokens (the throughput per instance in production is about tens or a few hundred of QPS). With higher hit rates (e.g., 100\% DRAM hit), \sysname sustains \textit{hundreds of QPS} even for sequences beyond 8K. Overall, \sysname expands the feasible operating region where long-sequence GR can be served under strict tail latency.

\paragraph{Latency composition under longer sequences.}
Figure~\ref{fig:exp2_2} reports the component latencies as sequence length increases. Pre-inference latency increases with sequence length, but it is smaller than baseline full inference because it only processes long-term behaviors (excluding candidate-item tokens). Both \texttt{load} and \texttt{rank} remain within tens of milliseconds even at large sequences, making them suitable for the ranking-stage P99 budget. In our setup, \sysname supports sequences up to $\sim$15K tokens with \texttt{load} (no concurrency) below 20\,ms and \texttt{rank} below 10\,ms (under the ranking-stage P99 SLO).

\paragraph{Cache loading under sequence growth and concurrency.}
Figure~\ref{fig:exp2_3} shows that DRAM$\rightarrow$HBM loading latency increases with both sequence length and concurrency, but remains far below baseline full inference. This aligns with the cost model: attention computation grows super-linearly with sequence length, whereas cache loading scales approximately linearly with cache size (about tens of megabytes per user). Thus, for long sequences, using stored $\psi$ to avoid recomputation is increasingly advantageous, especially at moderate concurrency and under tight SLO constraint.

\paragraph{Effect of retrieval slack.}
Figure~\ref{fig:exp2_4} studies how the retrieval-stage tail latency budget impacts the maximum supported concurrency. The baseline is unaffected because all computation happens in fine-grained ranking. In contrast, \sysname can exploit additional retrieval slack to perform more relay-race pre-inference, increasing the number of simultaneously supported queries. When retrieval P99 is 100\,ms, the maximum supported concurrency is about \textbf{5$\times$} that of the baseline in our experiments (more pre-inference are triggered during the retrieval). This suggests \sysname can trade retrieval slack for ranking tail-latency headroom in bursty scenarios.

\subsection{Q3: Extension of \sysname}
\label{sec:eval:rq3}

\paragraph{Sensitivity to candidate set size.}
Figure~\ref{fig:exp3_1} shows that \sysname’s ranking latency is primarily driven by the incremental input (short-term tokens and candidate items), and is substantially smaller than the baseline where long-prefix computation remains on the critical path. In production, candidate set size is typically around 512 per query; \sysname keeps ranking-on-cache latency well controlled (e.g., below 10\,ms even at 2048 items per ranking inference in our setup). This indicates the relay-race design decouples long-prefix cost from the per-query item scoring workload.

\paragraph{Utilization under concurrency.}
Figure~\ref{fig:exp3_2} reports NPU (cube) utilization under different concurrency levels. GR inference is compute-intensive; utilization increases with concurrency but also increases tail latency due to contention. \sysname with 0\% DRAM hit can increase utilization because it introduces additional pre-inference work; higher DRAM hit rates reduce this extra work and thus reduce utilization. Importantly, \sysname allows this utilization--latency trade-off to be tuned while keeping ranking-stage P99 intact.

\paragraph{Scaling to larger embedding dimensions.}
Figure~\ref{fig:exp3_3} shows that throughput decreases as embedding dimension increases for both baseline and \sysname, reflecting increased FLOPs. However, \sysname consistently provides higher SLO-compliant throughput: at 1024-dim (whose parameters are several times larger than that of the original HSTU), the baseline drops below 50 QPS in our setup, while \sysname achieves $\ge 2\times$ throughput, and \sysname with 100\% DRAM hit reaches $\sim 3\times$. This demonstrates that \sysname generalizes beyond sequence scaling to ``width'' scaling.

\paragraph{Scaling to deeper models.}
Figure~\ref{fig:exp3_4} shows that throughput decreases as model depth increases. When doubling the layers, baseline throughput drops by about 50\%, whereas \sysname exhibits a smoother degradation due to cached-prefix reuse; with 100\% hit, doubling layers reduces throughput by only about 14\%. When the backbone has 16 layers, \sysname achieves $\ge 4\times$ throughput over baseline, indicating relay-race inference remains beneficial for deeper GR blocks.

\begin{figure}[t]
    \begin{subfigure}[h]{0.45\textwidth}
        \includegraphics[width=0.8\textwidth]{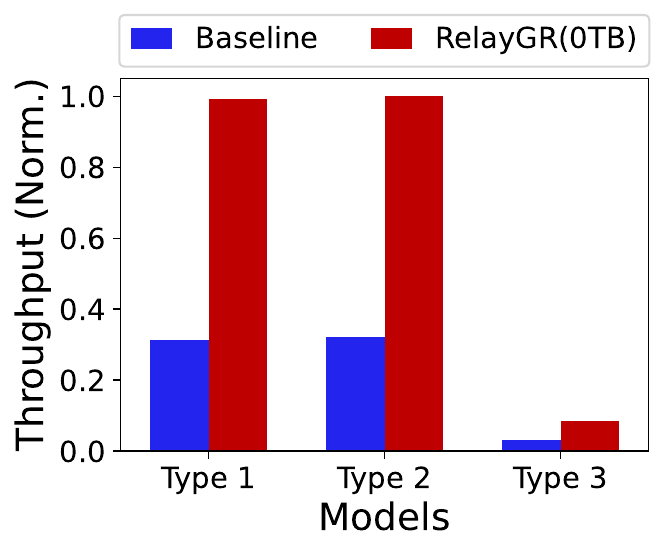}
        \caption{Various models} 
        \label{fig:exp_extra2_1}
    \end{subfigure}
    \begin{subfigure}[h]{0.45\textwidth}
        \includegraphics[width=0.8\textwidth]{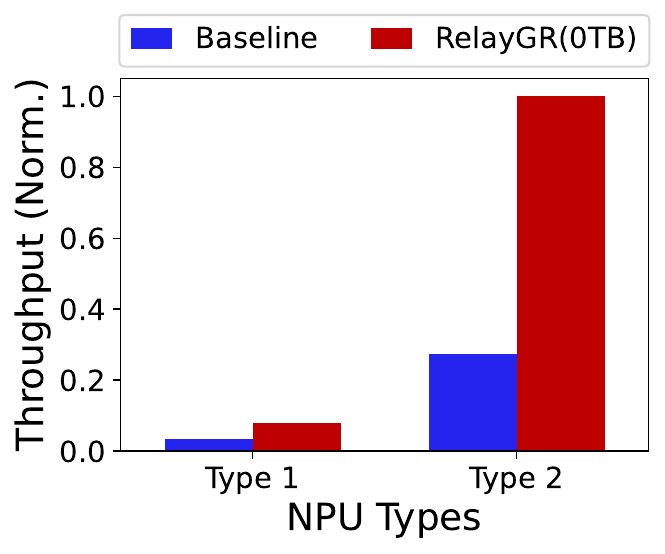}
        \caption{Various NPUs}
        \label{fig:exp_extra2_2}
    \end{subfigure}
    \caption{Generality across models and NPUs.}
    \label{fig:exp_extra2}
\end{figure}

\begin{table}[!t]
    \caption{KV caches under default settings.}
    \centering
    \vspace{-7pt}
    \begin{tabular}{c|cccc|c}
        \hline
        Model &Seq. &Layer &Format &Dim. &Size\\ \hline
        Type 1/2/3 &2K &8 &fp32 &256 &32MB\\ \hline
        \end{tabular}
    \vspace{-8pt}
    \label{table:cache_size}
\end{table}

\paragraph{Generality across models and NPUs.}
Finally, we evaluate \sysname across multiple GR models (Figure~\ref{fig:exp_extra2_1}) and Ascend NPU types (e.g., Ascend 310 (Type~1) and 910C (Type~2), Figure~\ref{fig:exp_extra2_2}). Although the absolute latency and throughput vary substantially across models—sometimes by nearly an order of magnitude under default settings—the relative trends are consistent: \sysname extends the maximum sequence length that can be served and increases SLO-compliant throughput under strict tail-latency constraints.
Model sizes also differ markedly. The combined Longer~\cite{chai2025longer} + RankMixer~\cite{zhu2025rankmixer} model (Type~3) is significantly larger than HSTU (Type~1) and its variant (Type~2). Under default settings, Table~\ref{table:cache_size} reports a per-request KV cache footprint of 32,MB. Note that Type~2 differs from Type~1 only in its attention computation, while for Type~3 we cache only the Longer component; RankMixer is implemented as a downstream DLRM in the fine-grained ranking stage.
Finally, even with a 2K-token input, the Type~1 baseline can exceed the P99 latency budget, making long-sequence GR infeasible without caching—further highlighting the practical value of \sysname.

%% file: tex/related.tex
\section{Related Work}

We discuss three lines of work most relevant to \sysname: (i) LLM serving architectures that manage KV caches across phases, (ii) generative recommendation (GR) models that motivate long-sequence inference, and (iii) production recommender optimizations.

\paragraph{LLM Serving: Prefill--Decode Placement and KVCache Management.}
A large body of LLM serving work studies how to place prefill and decode and how to manage KVCache to maximize goodput under SLOs. Representative systems disaggregate and coordinate prefill and decode (e.g., Splitwise~\cite{patel2024splitwise}, DistServe~\cite{zhong2024distserve}, TetriInfer~\cite{DBLP:journals/corr/abs-2401-11181}, P/D-Serve~\cite{jin2024pdserve}), scale deployment across GPU/NPU clusters (e.g., Mooncake~\cite{305212} and related production-scale deployments), dynamically balance resources between prefill and decode (e.g., DynaServe~\cite{ruan2025dynaserve}, Adrenaline~\cite{liang2025adrenaline}), and use distributed KVCache pools or cache-centric designs (e.g., MemServe~\cite{DBLP:journals/corr/abs-2406-17565}, CachedAttention~\cite{DBLP:conf/usenix/GaoHSKJDYYZ24}, TaiChi~\cite{wang2025taichi}).

Most of these systems operate on a two-stage inference graph: prefill produces a KVCache that decode consumes immediately. The producer therefore knows (or can decide) the consumer at handoff time, and cache movement is a direct phase-boundary operation. \sysname targets a different regime: GR caching is a \textit{lifecycle} problem across a multi-stage recommender pipeline. The prefix cache is produced early (during retrieval) and consumed later (during ranking) after pre-processing, while the eventual ranking instance is only determined after intermediate filtering. This setting introduces (i) a non-trivial survivability window, (ii) a placement problem under late-binding routing, and (iii) stringent P99 constraints under high QPS. \sysname addresses these challenges by jointly enforcing admission control (bounded live-cache footprint) and an affinity routing contract (no remote fetch on the ranking critical path). We thus study a distinct systems problem: \textit{lifecycle caching under late-binding placement}.

\paragraph{Generative Recommendation and Long-Sequence Modeling.}
Traditional recommender models (e.g., DLRMs and variants~\cite{DBLP:conf/kdd/WangFFW17,DBLP:conf/kdd/ZhouZSFZMYJLG18,DBLP:conf/aaai/ZhouMFPBZZG19,Huang_2019,chen2021eta,Wang_2021,zhu2025rankmixer}) have motivated extensive systems work on efficient embedding and ranking. More recently, GR models improve recommendation quality by modeling long sequential behaviors and exhibiting favorable scaling with longer sequences and larger capacity. Representative GR models include HSTU~\cite{zhai2024hstu}, LONGER~\cite{chai2025longer}, MTGR~\cite{Han_2025}, HLLM~\cite{chen2024hllm}, TIGER~\cite{DBLP:conf/nips/RajputMSKVHHT0S23}, LUM~\cite{yan2025lum}, Climber~\cite{Xu_2025}, GenRank~\cite{huang2025genrank}, OneRec~\cite{deng2025onerec}, OneRec-v2~\cite{zhou2025onerecv2}, OneRec-Think~\cite{liu2025onerecthink}, and OneTrans~\cite{zhang2025onetrans}. These works span discriminative usage for scoring/ranking (e.g., LONGER/MTGR/HLLM) and generative usage (e.g., producing embeddings or token-by-token targets, often with additional task towers).

Our work is complementary: we do not propose a new GR model. Instead, we address the \textit{online systems barrier} that prevents GR from realizing its offline scaling potential. Production ranking must satisfy tens-of-milliseconds tail latency; increasing sequence length or feature dimension inflates inference cost and quickly violates P99, forcing systems to cap online sequences. \sysname enables long-prefix reuse by pre-inferring the user's long-term behavior prefix and caching the per-layer KV, thereby removing expensive long-prefix computation from the ranking critical path. Compared with model-side compression (e.g., token compression), \sysname preserves long-sequence semantics while enforcing tail-latency constraints through admission control and placement.

\paragraph{Production Recommender Optimization.}
Industrial recommender stacks optimize end-to-end pipelines via kernel fusion, mixed precision, embedding caching, and efficient feature/embedding services. Tooling such as NVIDIA Merlin~\cite{merlin} and RecSys-oriented GR stacks~\cite{resys_gr} exemplify best practices for efficient training and inference. Other production efforts emphasize holistic pipeline optimization (e.g., co-design over the entire recommender)~\cite{DBLP:conf/nsdi/YangWYWDLZZLZWD25,fs_merlin,Wei_2022}.

These systems primarily optimize \textit{within-stage} execution (e.g., faster embedding lookup, fused operators, or more efficient model kernels). Such optimizations improve constant factors but do not fundamentally address the \textit{cross-stage} nature of long-sequence GR inference, where expensive prefix computation is separated from ranking by intermediate pipeline stages. \sysname is orthogonal: it introduces a cross-stage execution path that shifts long-prefix computation earlier and makes it reusable later under a lifecycle window and strict constraints. In particular, \sysname explicitly handles (i) survivability of prefix KV caches across pipeline stages, (ii) late-binding routing to the eventual ranking instance, and (iii) high-QPS contention, which are not the primary focus of within-stage kernel and embedding optimizations.
 

%% file: tex/conclusion.tex
\section{Conclusion}

Long-sequence generative recommendation is currently bottlenecked by the ranking-stage P99: as sequence length grows, full GR inference quickly becomes infeasible on the online critical path, preventing production systems from realizing the scaling gains observed offline. This paper shows that the key opportunity is to \textit{pre-infer} the user’s long-term behavior prefix and reuse its per-layer KV cache across the multi-stage recommender pipeline.

We presented \sysname, a production system that enables \textit{in-HBM relay-race inference}---a form of \textit{lifecycle caching under late-binding placement}---where prefix caches are produced early, survive through intermediate stages, and are consumed by ranking without remote fetches. \sysname combines (i) a sequence-aware trigger that admits only at-risk requests under bounded cache footprint and load, (ii) an affinity-aware router that co-locates cache production and consumption despite late-binding routing decisions, and (iii) a memory-aware expander that leverages server-local DRAM to extend reuse across repeated requests while avoiding redundant reloads.

We implemented \sysname on Ascend NPUs and evaluated it with real queries in a production-mirror environment. Under industrial P99 constraints, \sysname increases the maximum supported input sequence length by up to 1.5$\times$ and improves SLO-compliant throughput by up to 3.6$\times$, demonstrating that relay-race inference is an effective systems path to bringing long-sequence GR to real-time recommenders.

%% file: main.bib
@inproceedings {305212,
author = {Ruoyu Qin and Zheming Li and Weiran He and Jialei Cui and Feng Ren and Mingxing Zhang and Yongwei Wu and Weimin Zheng and Xinran Xu},
title = {Mooncake: Trading More Storage for Less Computation {\textemdash} A {KVCache-centric} Architecture for Serving {LLM} Chatbot},
booktitle = {23rd USENIX Conference on File and Storage Technologies (FAST 25)},
year = {2025}
}

@article{naumov2019dlrm,
    title   = {Deep Learning Recommendation Model for Personalization and Recommendation Systems},
    author  = {Maxim Naumov and Dheevatsa Mudigere and Hao-Jun Michael Shi and Jianyu Huang and Narayanan Sundaraman and Jongsoo Park and Xiaodong Wang and Udit Gupta and Carole-Jean Wu and Alisson G. Azzolini and Dmytro Dzhulgakov and Andrey Mallevich and Ilia Cherniavskii and Yinghai Lu and Raghuraman Krishnamoorthi and Ansha Yu and Volodymyr Kondratenko and Stephanie Pereira and Xianjie Chen and Wenlin Chen and Vijay Rao and Bill Jia and Liang Xiong and Misha Smelyanskiy},
    journal = {arXiv preprint arXiv:1906.00091},
    year    = {2019}
}

@article{DBLP:journals/corr/abs-2407-21022,
  author        = {Junjie Huang and
                  Jizheng Chen and
                  Jianghao Lin and
                  Jiarui Qin and
                  Ziming Feng and
                  Weinan Zhang and
                  Yong Yu},
  title         = {A Comprehensive Survey on Retrieval Methods in Recommender Systems},
  journal       = {arXiv preprint arXiv:2407.21022},
  year          = {2024}
}

@article{DBLP:journals/aim/BurkeFG11,
  author       = {Robin D. Burke and
                  Alexander Felfernig and
                  Mehmet H. G{\"{o}}ker},
  title        = {Recommender Systems: An Overview},
  journal      = {{AI} Mag.},
  volume       = {32},
  number       = {3},
  pages        = {13--18},
  year         = {2011},
}

@inproceedings{Wang_2021,
  title       = {DCN V2: Improved Deep {\&} Cross Network and Practical Lessons for Web-scale Learning to Rank Systems},
  booktitle   = {ACM WWW},
  author      = {Wang, Ruoxi and Shivanna, Rakesh and Cheng, Derek and Jain, Sagar and Lin, Dong and Hong, Lichan and Chi, Ed},
  pages       = {1785--1797},
  year        = {2021}
}

@inproceedings{Huang_2019, 
  title         = {FiBiNET: combining feature importance and bilinear feature interaction for click-through rate prediction},
  booktitle     = {ACM RecSys},
  author        = {Huang, Tongwen and Zhang, Zhiqi and Zhang, Junlin},
  pages         = {169--177},
  year          = {2019},
}

@article{zhu2025rankmixer,
    title   = {RankMixer: Scaling Up Ranking Models in Industrial Recommenders},
    author  = {Jie Zhu and Zhifang Fan and Xiaoxie Zhu and Yuchen Jiang and Hangyu Wang and Xintian Han and Haoran Ding and Xinmin Wang and Wenlin Zhao and Zhen Gong and Huizhi Yang and Zheng Chai and Zhe Chen and Yuchao Zheng and Qiwei Chen and Feng Zhang and Xun Zhou and Peng Xu and Xiao Yang and Di Wu and Zuotao Liu},
    journal = {arXiv preprint arXiv:2507.15551},
    year    = {2025}
}

@article{zhai2024hstu,
  title     = {Actions Speak Louder than Words: Trillion-Parameter Sequential Transducers for Generative Recommendations}, 
  author    = {Jiaqi Zhai and Lucy Liao and Xing Liu and Yueming Wang and Rui Li and Xuan Cao and Leon Gao and Zhaojie Gong and Fangda Gu and Michael He and Yinghai Lu and Yu Shi},
  journal   = {arXiv preprint arXiv:2402.17152},
  year      = {2024},
}

@inproceedings{Han_2025,
  title={MTGR: Industrial-Scale Generative Recommendation Framework in Meituan},
  booktitle={ACM CIKM},
  author={Han, Ruidong and Yin, Bin and Chen, Shangyu and Jiang, He and Jiang, Fei and Li, Xiang and Ma, Chi and Huang, Mincong and Li, Xiaoguang and Jing, Chunzhen and Han, Yueming and Zhou, MengLei and Yu, Lei and Liu, Chuan and Lin, Wei},
  year={2025},
}

@article{zhang2025onetrans,
  title     = {OneTrans: Unified Feature Interaction and Sequence Modeling with One Transformer in Industrial Recommender}, 
  author    = {Zhaoqi Zhang and Haolei Pei and Jun Guo and Tianyu Wang and Yufei Feng and Hui Sun and Shaowei Liu and Aixin Sun},
  journal   = {arXiv preprint arXiv:2510.26104},
  year      = {2025},
}

@article{deng2025onerec,
  title     = {OneRec: Unifying Retrieve and Rank with Generative Recommender and Iterative Preference Alignment}, 
  author    = {Jiaxin Deng and Shiyao Wang and Kuo Cai and Lejian Ren and Qigen Hu and Weifeng Ding and Qiang Luo and Guorui Zhou},
  journal   = {arXiv preprint arXiv:2502.18965},
  year      = {2025}, 
}

@article{zhou2025onerecv2,
  title     = {OneRec-V2 Technical Report}, 
  author    = {Guorui Zhou and Hengrui Hu and Hongtao Cheng and Huanjie Wang and Jiaxin Deng and Jinghao Zhang and Kuo Cai and Lejian Ren and Lu Ren and Liao Yu and Pengfei Zheng and Qiang Luo and Qianqian Wang and Qigen Hu and Rui Huang and Ruiming Tang and Shiyao Wang and Shujie Yang and Tao Wu and Wuchao Li and Xinchen Luo and Xingmei Wang and Yi Su and Yunfan Wu and Zexuan Cheng and Zhanyu Liu and Zixing Zhang and Bin Zhang and Boxuan Wang and Chaoyi Ma and Chengru Song and Chenhui Wang and Chenglong Chu and Di Wang and Dongxue Meng and Dunju Zang and Fan Yang and Fangyu Zhang and Feng Jiang and Fuxing Zhang and Gang Wang and Guowang Zhang and Han Li and Honghui Bao and Hongyang Cao and Jiaming Huang and Jiapeng Chen and Jiaqiang Liu and Jinghui Jia and Kun Gai and Lantao Hu and Liang Zeng and Qiang Wang and Qidong Zhou and Rongzhou Zhang and Shengzhe Wang and Shihui He and Shuang Yang and Siyang Mao and Sui Huang and Tiantian He and Tingting Gao and Wei Yuan and Xiao Liang and Xiaoxiao Xu and Xugang Liu and Yan Wang and Yang Zhou and Yi Wang and Yiwu Liu and Yue Song and Yufei Zhang and Yunfeng Zhao and Zhixin Ling and Ziming Li},
  journal   = {arXiv preprint arXiv:2508.20900},
  year      = {2025}, 
}

@article{liu2025onerecthink,
  title     = {OneRec-Think: In-Text Reasoning for Generative Recommendation}, 
  author    = {Zhanyu Liu and Shiyao Wang and Xingmei Wang and Rongzhou Zhang and Jiaxin Deng and Honghui Bao and Jinghao Zhang and Wuchao Li and Pengfei Zheng and Xiangyu Wu and Yifei Hu and Qigen Hu and Xinchen Luo and Lejian Ren and Zixing Zhang and Qianqian Wang and Kuo Cai and Yunfan Wu and Hongtao Cheng and Zexuan Cheng and Lu Ren and Huanjie Wang and Yi Su and Ruiming Tang and Kun Gai and Guorui Zhou},
  journal   = {arXiv preprint arXiv:2510.11639},
  year      = {2025},
}

@article{rajput2023tiger,
  title     = {Recommender Systems with Generative Retrieval}, 
  author    = {Shashank Rajput and Nikhil Mehta and Anima Singh and Raghunandan H. Keshavan and Trung Vu and Lukasz Heldt and Lichan Hong and Yi Tay and Vinh Q. Tran and Jonah Samost and Maciej Kula and Ed H. Chi and Maheswaran Sathiamoorthy},
  journal   = {arXiv preprint arXiv:2305.05065},
  year      = {2023},
}

@article{yang2025grllm,
  title     = {GR-LLMs: Recent Advances in Generative Recommendation Based on Large Language Models}, 
  author    = {Zhen Yang and Haitao Lin and Jiawei xue and Ziji Zhang},
  journal   = {arXiv preprint arXiv:2507.06507},
  year      = {2025}, 
}

@inproceedings{DBLP:conf/ijcai/0001LGQZ0T23,
  author       = {Zhicheng He and
                  Weiwen Liu and
                  Wei Guo and
                  Jiarui Qin and
                  Yingxue Zhang and
                  Yaochen Hu and
                  Ruiming Tang},
  title        = {A Survey on User Behavior Modeling in Recommender Systems},
  booktitle    = {{IJCAI}},
  pages        = {6656--6664},
  year         = {2023},
}

@article{patel2024splitwise,
  title     = {Splitwise: Efficient generative LLM inference using phase splitting}, 
  author    = {Pratyush Patel and Esha Choukse and Chaojie Zhang and Aashaka Shah and Íñigo Goiri and Saeed Maleki and Ricardo Bianchini},
  journal   = {arXiv preprint arXiv:2311.18677},
  year      = {2024},
}

@article{zhong2024distserve,
  title     = {DistServe: Disaggregating Prefill and Decoding for Goodput-optimized Large Language Model Serving}, 
  author    = {Yinmin Zhong and Shengyu Liu and Junda Chen and Jianbo Hu and Yibo Zhu and Xuanzhe Liu and Xin Jin and Hao Zhang},
  journal   = {arXiv preprint arXiv:2401.09670},
  year      = {2024},
}

@article{jin2024pdserve,
  title     = {P/D-Serve: Serving Disaggregated Large Language Model at Scale}, 
  author    = {Yibo Jin and Tao Wang and Huimin Lin and Mingyang Song and Peiyang Li and Yipeng Ma and Yicheng Shan and Zhengfan Yuan and Cailong Li and Yajing Sun and Tiandeng Wu and Xing Chu and Ruizhi Huan and Li Ma and Xiao You and Wenting Zhou and Yunpeng Ye and Wen Liu and Xiangkun Xu and Yongsheng Zhang and Tiantian Dong and Jiawei Zhu and Zhe Wang and Xijian Ju and Jianxun Song and Haoliang Cheng and Xiaojing Li and Jiandong Ding and Hefei Guo and Zhengyong Zhang},
  journal   = {arXiv preprint arXiv:2408.08147},
  year      = {2024},
}

@inproceedings{DBLP:conf/www/EksombatchaiJLL18,
  author       = {Chantat Eksombatchai and
                  Pranav Jindal and
                  Jerry Zitao Liu and
                  Yuchen Liu and
                  Rahul Sharma and
                  Charles Sugnet and
                  Mark Ulrich and
                  Jure Leskovec},
  title        = {Pixie: {A} System for Recommending 3+ Billion Items to 200+ Million
                  Users in Real-Time},
  booktitle    = {{ACM} {WWW}},
  pages        = {1775--1784},
  year         = {2018},
}

@inproceedings{DBLP:conf/recsys/WangLLMWGZHBBCC24,
  author       = {Jianling Wang and
                  Haokai Lu and
                  Yifan Liu and
                  He Ma and
                  Yueqi Wang and
                  Yang Gu and
                  Shuzhou Zhang and
                  Ningren Han and
                  Shuchao Bi and
                  Lexi Baugher and
                  Ed H. Chi and
                  Minmin Chen},
  title        = {LLMs for User Interest Exploration in Large-scale Recommendation Systems},
  booktitle    = {{ACM} RecSys},
  pages        = {872--877},
  year         = {2024},
}

@article{DBLP:journals/corr/abs-2401-11181,
  author       = {Cunchen Hu and
                  Heyang Huang and
                  Liangliang Xu and
                  Xusheng Chen and
                  Jiang Xu and
                  Shuang Chen and
                  Hao Feng and
                  Chenxi Wang and
                  Sa Wang and
                  Yungang Bao and
                  Ninghui Sun and
                  Yizhou Shan},
  title        = {Inference without Interference: Disaggregate {LLM} Inference for Mixed Downstream Workloads},
  journal      = {arXiv preprint arXiv:2401.11181},
  year         = {2024},
}

@article{DBLP:journals/corr/abs-2406-17565,
  author       = {Cunchen Hu and
                  Heyang Huang and
                  Junhao Hu and
                  Jiang Xu and
                  Xusheng Chen and
                  Tao Xie and
                  Chenxi Wang and
                  Sa Wang and
                  Yungang Bao and
                  Ninghui Sun and
                  Yizhou Shan},
  title        = {MemServe: Context Caching for Disaggregated {LLM} Serving with Elastic Memory Pool},
  journal      = {arXiv preprint arXiv:2406.17565},
  year         = {2024},
}

@article{ruan2025dynaserve,
  title     = {DynaServe: Unified and Elastic Execution for Dynamic Disaggregated LLM Serving}, 
  author    = {Chaoyi Ruan and Yinhe Chen and Dongqi Tian and Yandong Shi and Yongji Wu and Jialin Li and Cheng Li},
  journal   = {arXiv preprint arXiv:2504.09285},
  year      = {2025}, 
}

@inproceedings{DBLP:conf/usenix/GaoHSKJDYYZ24,
  author       = {Bin Gao and
                  Zhuomin He and
                  Puru Sharma and
                  Qingxuan Kang and
                  Djordje Jevdjic and
                  Junbo Deng and
                  Xingkun Yang and
                  Zhou Yu and
                  Pengfei Zuo},
  title        = {Cost-Efficient Large Language Model Serving for Multi-turn Conversations with CachedAttention},
  booktitle    = {{USENIX} {ATC}},
  pages        = {111--126},
  year         = {2024},
}

@article{liang2025adrenaline,
  title     = {Injecting Adrenaline into LLM Serving: Boosting Resource Utilization and Throughput via Attention Disaggregation}, 
  author    = {Yunkai Liang and Zhangyu Chen and Pengfei Zuo and Zhi Zhou and Xu Chen and Zhou Yu},
  journal   = {arXiv preprint arXiv:2503.20552},
  year      = {2025},
}

@article{wang2025taichi,
  title     = {Prefill-Decode Aggregation or Disaggregation? Unifying Both for Goodput-Optimized LLM Serving}, 
  author    = {Chao Wang and Pengfei Zuo and Zhangyu Chen and Yunkai Liang and Zhou Yu and Ming-Chang Yang},
  journal   = {arXiv preprint arXiv:2508.01989},
  year      = {2025},
}

@inproceedings{DBLP:conf/kdd/ZhouZSFZMYJLG18,
  author       = {Guorui Zhou and
                  Xiaoqiang Zhu and
                  Chengru Song and
                  Ying Fan and
                  Han Zhu and
                  Xiao Ma and
                  Yanghui Yan and
                  Junqi Jin and
                  Han Li and
                  Kun Gai},
  title        = {Deep Interest Network for Click-Through Rate Prediction},
  booktitle    = {{ACM} {SIGKDD}},
  pages        = {1059--1068},
  year         = {2018},
}

@inproceedings{DBLP:conf/aaai/ZhouMFPBZZG19,
  author       = {Guorui Zhou and
                  Na Mou and
                  Ying Fan and
                  Qi Pi and
                  Weijie Bian and
                  Chang Zhou and
                  Xiaoqiang Zhu and
                  Kun Gai},
  title        = {Deep Interest Evolution Network for Click-Through Rate Prediction},
  booktitle    = {{AAAI}},
  pages        = {5941--5948},
  year         = {2019},
}

@inproceedings{DBLP:conf/kdd/WangFFW17,
  author       = {Ruoxi Wang and
                  Bin Fu and
                  Gang Fu and
                  Mingliang Wang},
  title        = {Deep {\&} Cross Network for Ad Click Predictions},
  booktitle    = {{ACM} {ADKDD}},
  pages        = {12:1--12:7},
  year         = {2017},
}

@article{chen2021eta,
  title     = {End-to-End User Behavior Retrieval in Click-Through Rate Prediction Model}, 
  author    = {Qiwei Chen and Changhua Pei and Shanshan Lv and Chao Li and Junfeng Ge and Wenwu Ou},
  journal   = {arXiv preprint arXiv:2108.04468},
  year      = {2021},
}

@article{chai2025longer,
  title     = {LONGER: Scaling Up Long Sequence Modeling in Industrial Recommenders}, 
  author    = {Zheng Chai and Qin Ren and Xijun Xiao and Huizhi Yang and Bo Han and Sijun Zhang and Di Chen and Hui Lu and Wenlin Zhao and Lele Yu and Xionghang Xie and Shiru Ren and Xiang Sun and Yaocheng Tan and Peng Xu and Yuchao Zheng and Di Wu},
  journal   = {arXiv preprint arXiv:2505.04421},
  year      = {2025},
}

@article{chen2024hllm,
  title     = {HLLM: Enhancing Sequential Recommendations via Hierarchical Large Language Models for Item and User Modeling}, 
  author    = {Junyi Chen and Lu Chi and Bingyue Peng and Zehuan Yuan},
  journal   = {arXiv preprint arXiv:2409.12740},
  year      = {2024},
}

@inproceedings{DBLP:conf/nips/RajputMSKVHHT0S23,
  author       = {Shashank Rajput and
                  Nikhil Mehta and
                  Anima Singh and
                  Raghunandan Hulikal Keshavan and
                  Trung Vu and
                  Lukasz Heldt and
                  Lichan Hong and
                  Yi Tay and
                  Vinh Q. Tran and
                  Jonah Samost and
                  Maciej Kula and
                  Ed H. Chi and
                  Mahesh Sathiamoorthy},
  title        = {Recommender Systems with Generative Retrieval},
  booktitle    = {{NeurIPS}},
  year         = {2023},
}

@article{yan2025lum,
  title     = {Unlocking Scaling Law in Industrial Recommendation Systems with a Three-step Paradigm based Large User Model}, 
  author    = {Bencheng Yan and Shilei Liu and Zhiyuan Zeng and Zihao Wang and Yizhen Zhang and Yujin Yuan and Langming Liu and Jiaqi Liu and Di Wang and Wenbo Su and Wang Pengjie and Jian Xu and Bo Zheng},
  journal   = {arXiv preprint arXiv:2502.08309},
  year      = {2025},
}

@inproceedings{Xu_2025,
  title     = {Climber: Toward Efficient Scaling Laws for Large Recommendation Models},
  booktitle = {{ACM} {CIKM}},
  author    = {Xu, Songpei and Wang, Shijia and Guo, Da and Guo, Xianwen and Xiao, Qiang and Huang, Bin and Wu, Guanlin and Luo, Chuanjiang},
  year      = {2025},
  pages     = {6193–6200},
}

@article{huang2025genrank,
  title     = {Towards Large-scale Generative Ranking}, 
  author    = {Yanhua Huang and Yuqi Chen and Xiong Cao and Rui Yang and Mingliang Qi and Yinghao Zhu and Qingchang Han and Yaowei Liu and Zhaoyu Liu and Xuefeng Yao and Yuting Jia and Leilei Ma and Yinqi Zhang and Taoyu Zhu and Liujie Zhang and Lei Chen and Weihang Chen and Min Zhu and Ruiwen Xu and Lei Zhang},
  journal   = {arXiv preprint arXiv:2505.04180},
  year      = {2025},
}

@article{skrlj2025dcn2,
  title     = {DCN{\^{}}2: Interplay of Implicit Collision Weights and Explicit Cross Layers for Large-Scale Recommendation}, 
  author       = {Blaz Skrlj and
                  Yonatan Karni and
                  Grega Gaspersic and
                  Blaz Mramor and
                  Yulia Stolin and
                  Martin Jakomin and
                  Jasna Urbancic and
                  Yuval Dishi and
                  Natalia Silberstein and
                  Ophir Friedler and
                  Assaf Klein},
  journal   = {arXiv preprint arXiv:2506.21624},
  year      = {2025},
}

@inproceedings{DBLP:conf/wsdm/ChenCXC21,
  author       = {Minmin Chen and
                  Bo Chang and
                  Can Xu and
                  Ed H. Chi},
  title        = {User Response Models to Improve a {REINFORCE} Recommender System},
  booktitle    = {{ACM} {WSDM}},
  pages        = {121--129},
  year         = {2021},
}

@inproceedings{DBLP:conf/nsdi/YangWYWDLZZLZWD25,
  author       = {Lingyun Yang and
                  Yongchen Wang and
                  Yinghao Yu and
                  Qizhen Weng and
                  Jianbo Dong and
                  Kan Liu and
                  Chi Zhang and
                  Yanyi Zi and
                  Hao Li and
                  Zechao Zhang and
                  Nan Wang and
                  Yu Dong and
                  Menglei Zheng and
                  Lanlan Xi and
                  Xiaowei Lu and
                  Liang Ye and
                  Guodong Yang and
                  Binzhang Fu and
                  Tao Lan and
                  Liping Zhang and
                  Lin Qu and
                  Wei Wang},
  title        = {GPU-Disaggregated Serving for Deep Learning Recommendation Models at Scale},
  booktitle    = {{USENIX} {NSDI}},
  pages        = {847--863},
  year         = {2025},
}

@inproceedings{DBLP:conf/osdi/LaiZLTWHDHPLCWR23,
  author       = {Fan Lai and
                  Wei Zhang and
                  Rui Liu and
                  William Tsai and
                  Xiaohan Wei and
                  Yuxi Hu and
                  Sabin Devkota and
                  Jianyu Huang and
                  Jongsoo Park and
                  Xing Liu and
                  Zeliang Chen and
                  Ellie Wen and
                  Paul Rivera and
                  Jie You and
                  Chun{-}cheng Jason Chen and
                  Mosharaf Chowdhury},
  title        = {AdaEmbed: Adaptive Embedding for Large-Scale Recommendation Models},
  booktitle    = {{USENIX} {OSDI}},
  pages        = {817--831},
  year         = {2023},
}

@article{zuo2025cloudmatrix,
  title     = {Serving Large Language Models on Huawei CloudMatrix384}, 
  author    = {Pengfei Zuo and Huimin Lin and Junbo Deng and Nan Zou and Xingkun Yang and Yingyu Diao and Weifeng Gao and Ke Xu and Zhangyu Chen and Shirui Lu and Zhao Qiu and Peiyang Li and Xianyu Chang and Zhengzhong Yu and Fangzheng Miao and Jia Zheng and Ying Li and Yuan Feng and Bei Wang and Zaijian Zong and Mosong Zhou and Wenli Zhou and Houjiang Chen and Xingyu Liao and Yipeng Li and Wenxiao Zhang and Ping Zhu and Yinggang Wang and Chuanjie Xiao and Depeng Liang and Dong Cao and Juncheng Liu and Yongqiang Yang and Xiaolong Bai and Yi Li and Huaguo Xie and Huatao Wu and Zhibin Yu and Lv Chen and Hu Liu and Yujun Ding and Haipei Zhu and Jing Xia and Yi Xiong and Zhou Yu and Heng Liao},
  journal   = {arXiv preprint arXiv:2506.12708},
  year      = {2025},
}

@misc{fs_merlin,
  title        = {{Offline to Online: Feature Storage for Real-time Recommendation Systems with NVIDIA Merlin}},
  author       = {{Nvidia Technical Blog}},
  year         = {2023},
  howpublished = {\url{https://developer.nvidia.com/blog/offline-to-online-feature-storage-for-real-time-recommendation-systems-with-nvidia-merlin/}}
}

@misc{merlin,
  title        = {{NVIDIA-Merlin}},
  author       = {{NVIDIA Developers}},
  year         = {2025},
  howpublished = {\url{https://github.com/NVIDIA-Merlin/Merlin}}
}

@inproceedings{on_device_reranking,
  author = {Xi, Yunjia and Liu, Weiwen and Wang, Yang and Tang, Ruiming and Zhang, Weinan and Zhu, Yue and Zhang, Rui and Yu, Yong},
  title = {On-device Integrated Re-ranking with Heterogeneous Behavior Modeling},
  year = {2023},
  booktitle = {{ACM} {SIGKDD}},
  pages = {5225–5236},
}

@article{xi2025seral,
  title     = {Bursting Filter Bubble: Enhancing Serendipity Recommendations with Aligned Large Language Models}, 
  author    = {Yunjia Xi and Muyan Weng and Wen Chen and Chao Yi and Dian Chen and Gaoyang Guo and Mao Zhang and Jian Wu and Yuning Jiang and Qingwen Liu and Yong Yu and Weinan Zhang},
  journal   = {arXiv preprint arXiv:2502.13539},
  year      = {2025},
}

@article{gim2024promptcache,
  title     = {Prompt Cache: Modular Attention Reuse for Low-Latency Inference}, 
  author    = {In Gim and Guojun Chen and Seung-seob Lee and Nikhil Sarda and Anurag Khandelwal and Lin Zhong},
  journal   = {arXiv preprint arXiv:2311.04934},
  year      = {2024},
}

@article{ye2024chunkattention,
  title     = {ChunkAttention: Efficient Self-Attention with Prefix-Aware KV Cache and Two-Phase Partition}, 
  author    = {Lu Ye and Ze Tao and Yong Huang and Yang Li},
  journal   = {arXiv preprint arXiv:2402.15220},
  year      = {2024},
}

@inproceedings{10.1145/3725783.3764389,
  author = {Huang, Yibo and Qiu, Yiming and Yang, Zhenning and Dai, Yi and Wu, Dingming and Lai, Fan and Xing, Jiarong and Chen, Ang},
  title = {Towards Fully Disaggregated Recommendation Model Serving},
  year = {2025},
  booktitle = {{ACM} {SIGOPS} {APSys}},
  pages = {38–45},
}

@inproceedings{DBLP:conf/eurosys/XieLLWGRS22,
  author       = {Minhui Xie and
                  Youyou Lu and
                  Jiazhen Lin and
                  Qing Wang and
                  Jian Gao and
                  Kai Ren and
                  Jiwu Shu},
  title        = {Fleche: an efficient {GPU} embedding cache for personalized recommendations},
  booktitle    = {{ACM} {EuroSys}},
  pages        = {402--416},
  year         = {2022},
}

@misc{resys_gr,
  title        = {{NVIDIA RecSys Examples}},
  author       = {{NVIDIA}},
  year         = {2025},
  howpublished = {\url{https://github.com/NVIDIA/recsys-examples}}
}

@inproceedings{Wei_2022,
  title     = {A GPU-specialized Inference Parameter Server for Large-Scale Deep Recommendation Models},
  booktitle = {{ACM} {RecSys}},
  author    = {Wei, Yingcan and Langer, Matthias and Yu, Fan and Lee, Minseok and Liu, Jie and Shi, Ji and Wang, Zehuan},
  year      = {2022},
  pages     = {408–419},
}
